\pdfoutput=1

\documentclass[aps,prl,reprint,twocolumn,preprintnumbers,superscriptaddress,letterpaper]{revtex4-2}
\usepackage[english]{babel}
\usepackage{ucs}
\usepackage[utf8x]{inputenc}
\usepackage{amsmath}
\usepackage{amsfonts}
\usepackage{amssymb}
\usepackage{graphicx}
\usepackage{bm}
\usepackage{bbm}
\usepackage{empheq}
\usepackage{xr-hyper}
\usepackage{hyperref}
\usepackage{xcolor}

\newcommand{\ea}{\emph{et al.}}

\newcommand{\rb}{\overline{r}}

\begin{document}
\title{Optimal design of a superconducting transmon qubit with tapered wiring}
\author{John M. Martinis}
\affiliation{Quantala, Santa Barbara, Ca 93105, USA}
\email{martinis@quantala.tech}
\date{\today}

\begin{abstract}
Analytical formulas are presented for simplified but useful qubit geometries that predict surface dielectric loss when its thickness is much less than the metal thickness, the limiting case needed for real devices.  These formulas can thus be used to precisely predict loss and optimize the qubit layout.  Surprisingly, a significant fraction of surface loss comes from the small wire that connects the Josephson junction to the qubit capacitor.   Tapering this wire is shown to significantly lower its loss.  Also predicted are the size and density of the two-level state (TLS) spectrum from individual surface dissipation sites.   
\end{abstract}
\maketitle


Quantum computers are made from quantum bits, which have natural sources of noise and dissipation that produce errors in quantum gates.  Decreasing these errors increases the size and complexity of quantum algorithms that can be run on a quantum computer.  When errors are reduced to about 0.1\% per gate operation, then quantum error correction may be used on a large array of qubits in order to lower logical errors and execute vastly more complex quantum algorithms \cite{surfcode}.  Qubit errors are often limited by the rate of energy decay from loss mechanisms.
 
\begin{figure}[b]
\includegraphics[width=0.48\textwidth, 
trim = 100 200 110 70,clip]
{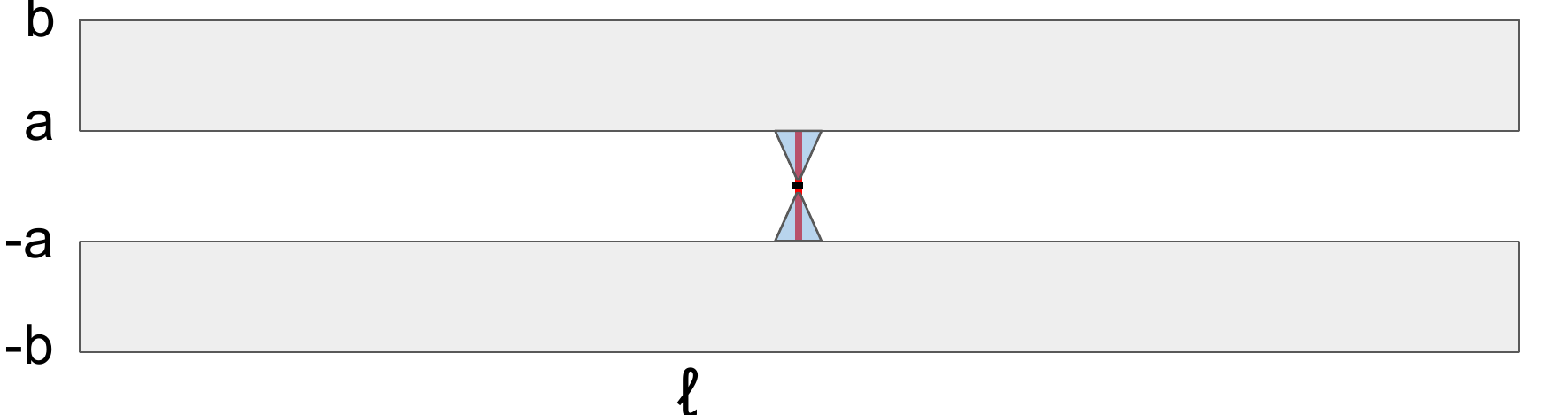}
\caption{\textbf{Qubit design with taper.}
Drawing of prototypical qubit device, with relative dimensions approximately to scale.  The qubit capacitance $C \simeq 100\,\textrm{fF}$ is made from two ribbons (shaded gray) of width $b-a = 100\,\mu\textrm{m}$ and length $\ell = 1300\,\mu\textrm{m}$, separated by distance $2a = 100\,\mu\textrm{m}$. Wires (drawn in red) connect to the sub-$\mu$m Josephson junction in a conventional design.  Surprisingly, the loss from the small junction wires is about equal to that coming from the large ribbons.  This paper proposes tapering these wires (shaded blue) to reduce their loss.  
}
\label{fig:Qubit}
\end{figure}

Superconducting qubits can be thought of as an inductor-capacitor resonator, with the superconducting Josephson junction giving a non-linear inductance that allows the two lowest energy levels to behave as a qubit.  The Josephson junction and the capacitance are designed to be separate physical entities, as illustrated in Fig.\,\ref{fig:Qubit}, and thus can be separately optimized..  The size of the Josephson junction is about 100\,nm.  Its natural capacitance is negligible and junction defects are statistically unlikely because of its small size; the junction can thus typically be modeled as bringing no energy loss.  The capacitance is made from superconducting pads with a relatively large millimeter size and about $100\,\mu\textrm{m}$ spacing, producing a capacitance of about 100\,fF for the transmon qubit \cite{transmon}.  When the capacitor is designed properly with control lines weakly coupled to an external circuit, dielectric surface loss from the superconductor and substrate is the dominant mechanism of energy loss.  As for any surface loss mechanism, it has been found experimentally that increasing the size of this capacitor lowers the net effect of the surface loss on the qubit device \cite{Yale}.

Calculating the surface loss is difficult because of the divergence of the electric fields at the metal edges, which has pushed researchers to solve the problem numerically with finite-element models \cite{wenner_surfloss,sandberg,wang,calusine,gambetta}.  More recently, an analytical result was obtained using solutions of conformal mapping, which describes the electric fields of ribbon and coplanar geometries \cite{IBM}.  Unfortunately, this result is only approximate since it assumed the dielectric is thicker than the metal, opposite of the real design.  Here, a more practical solution is presented that is valid for a few nanometer lossy dielectric surrounding a much thicker metal layer about $0.1\,\mu$m.  Changes to the conformal predictions are calculated using the scaling of corner fields and numerical simulation, and are simple to use and understand.

For planar transmons, ribbon capacitors are typically embedded in a ground plane, and thus the analytical results are not valid.  Numerical simulations and fit functions give capacitance and surface loss for this important practical case.   

New surface loss predictions are also given for the wires that connect the Josephson junction to the capacitor pads.  They typically have a width approximately the size of the junction, about $0.1\,\mu$m, and extend in length from the junction to the pads, about $50\,\mu\textrm{m}$.  They are typically narrow compared to their length and thus have large electric fields at its edges.  Conformal and numerical solutions are used to give analytic predictions of surface loss.  Their long length produces significant surface loss, which can be reduced by tapering the wires, as as illustrated in Fig.\,\ref{fig:Qubit}.  

For numerical solutions, meshing is always a concern given the range of size scales, from nanometer thick oxides to millimeter sized capacitor pads.  Meshing is particularly important for 2-D and 3-D numerical solvers where the large grid makes it more difficult to calculate edge fields accurately.  In this paper, the assumption of flat substrates (no trenching) allows solutions based on surface charges that are effectively 1-D, so that fine meshing at corners enables accurate checking of formulas.  These formulas are thus a useful ``gold standard'' reference for verifying numerical methods.  This is especially needed for experiments where surface loss parameters want to be accurately extracted \cite{partLL1,partLL}.  

Expressing loss with formulas is also useful since the designer can separate out all the loss mechanisms, instead of modeling the entire device at once using numerical solvers.  Optimization is more transparent, for example trading off the surface loss of the qubit capacitance pads and the junction wires.  In order to give useful design formulas for various geometries of the qubit capacitor, the surface loss is analyzed for 3 cases: a parallel plate, a ribbon capacitor where electric fields are between the two electrodes, and a coplanar capacitor where the fields connect through a ground plane.  A typical design should be able to be modeled as a combination of these geometries, thus enabling surface loss predictions from formulas derived here.

\section{Qubit Model and Participation Ratios}

\begin{figure}[t]
\includegraphics[width=0.48\textwidth, 
trim = 140 20 160 30,clip]
{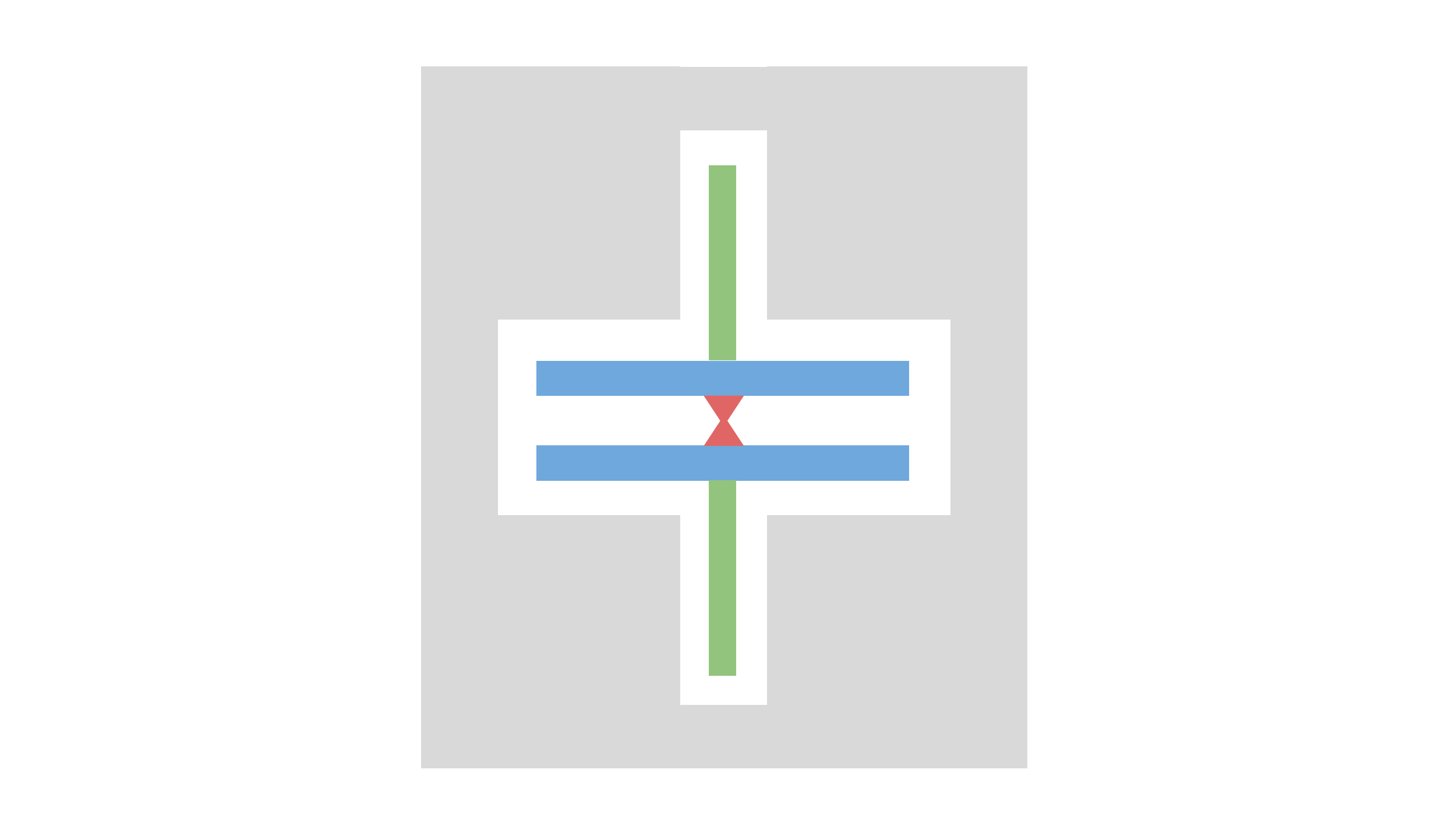}
\caption{\textbf{Example of full transmon design.}
Example design of a differential transmon qubit that incorporates several of the capacitance structures described here.  Red shows the junction and tapered wire, blue is a ribbon capacitor, and green is a coplanar capacitor.  The outer ground plan is gray.  The capacitances and losses would be added to give a good approximation for the entire design.    
}
\label{fig:full}
\end{figure}

Figure\,\ref{fig:full} shows an example design of a full differential qubit, include the qubit electrodes and an shielding ground plane (gray).  The full design can be broken up into the junction and tapered wires (red), a ribbon capacitor (blue) and a coplanar capacitor (green).  The capacitances and losses from the combination of the structures would then be used to optimize the design. 

We are interested here in calculating the loss from dielectrics, since the loss from the metallic structures are typically negligible for superconductors.  For a crystalline substrate  such as silicon or sapphire, the loss is dominated by the thin surface layers of the films \cite{wenner_surfloss}: the metal-air (MA), metal-substrate (MS) and substrate-air (SA), typically coming from amorphous oxides.

The total loss tangent for these thin layers is given by $\Sigma p_i \tan \delta_i$,  where surface interface type $i$ has loss tangent $\tan \delta_i$
and participation ratio of the stored energy
\begin{align}
    p_i = \frac{\epsilon_i/2}{W} \int dA \, t_i\,|E_i|^2
\end{align}
where the normal volume integral is replaced by a surface integral $dA$ for a thin dielectric layer with thickness $t_i$, dielectric constant $\epsilon_i$, and a surface electric field $E_i$.  The participation ratio is normalized by the total total capacitor energy $W = CV^2/2$, where $C$ is the total capacitance and $V$ the voltage.

When designing the qubit, the qubit capacitance $C$ is usually fixed to a desired parameter.  Because the results are more easy to interpret in terms of design distances, it is convenient to describe the qubit capacitance in terms of a length using
\begin{align}
    C \equiv \epsilon_0 L \ .
\end{align}
For $C = 100\,\textrm{fF}$, a value used for a qubit non-linearity of about 200\,MHz, one finds $L = 11.3\,\textrm{mm}$.  
 
For thin films, the electric fields of the top and bottom dielectrics can be considered separately.  Thus the electric fields $E_0$ can be solved for $\epsilon = \epsilon_0$, the free space value, and then multiplied by 1 for the solution on the air side, and by $\epsilon_s$ for the substrate side.  The surface dielectrics can be taken into account with the three participation ratios \cite{wenner_surfloss}
\begin{align}
    p_\textrm{MA} =& 
    \frac{1}{\epsilon_\textrm{MA}}
    \frac{t_\textrm{MA}}{L\epsilon_0 /2} 
    \Big[\frac{\epsilon_0}{2}\int_\textrm{MA} dA \, \,|E_0/V|^2 \Big] \ , \label{eq:MA} \\
    p_\textrm{MS} =&
    \frac{\epsilon_s^2}{\epsilon_\textrm{MS}} \,
    \frac{t_\textrm{MS}}{L\epsilon_0 /2}
    \Big[ \frac{\epsilon_0}{2} \int_\textrm{MS} dA \, \,|E_0/V|^2 \Big]  \ , \label{eq:MS}\\
    p_\textrm{SA} =& \ 
    \epsilon_\textrm{SA} \,
    \frac{t_\textrm{SA}}{L \epsilon_0 /2} 
    \Big[ \frac{\epsilon_0}{2} \int_\textrm{SA}  dA\, \,|E_0/V|^2 \Big] \ ,  \label{eq:SA}
\end{align}
where the area integrals correspond to the appropriate surfaces for each type, and the bracketed terms are called surface energies.  Tangential fields are only included for the SA formula, as appropriate for thin films \cite{wenner_surfloss}.  For the above MA and MS surface energies, the electric field is for only one side of the metal.  Thus for the total surface energy $U$ calculated in the next sections, the above MA and MS energies in brackets should be $U/2$. 

Dielectric constants are for a silicon substrate $\epsilon_s = 11.7$, aluminum oxide $\epsilon_\textrm{MA}=\epsilon_\textrm{MS} = 9.8$, and silicon dioxide $\epsilon_{SA} = 3.8$; the relative weights of the MA:MS:SA dielectric terms are $0.10:14:3.8$.  

\section{Differential Parallel Plate Capacitor}

The simplest geometry is a parallel plate capacitor.  This contribution is typically needed when transmon qubits are made using bump-bonded substrates, where the second substrate acts as a ground plane above the qubit metal pads and thus adds capacitance to the qubit.  This structure can be treated as parallel plate with each plate having width $w$ and length $\ell$ and a separation $s$ to the ground plane, with capacitance
\begin{align}
    C_\textrm{p} = (1/2) \epsilon_0 \ell w/s \ ,
\end{align}
where the $1/2$ factor coming from the differential design of the qubit, where the capacitance of each parallel plate is in series.   

The electric field in each differentially-driven capacitor is $E_\textrm{p} = \tfrac{1}{2}V/s$, and the total surface energy is
\begin{align}
        U_\textrm{p} & = (\epsilon/2) 2(2\ell w)(V/2s)^2 \ ,
\end{align}
where a factor of 2 comes from the two parallel plates, and another from surface loss at the 2 plates of the capacitor.  

The participation ratio for the parallel-plate capacitor is
\begin{align}
    p_\textrm{MA}^\textrm{p} =&  
    \frac{1}{\epsilon_\textrm{MA}} 
    \frac{t_\textrm{MA}}{L}
    \frac{\ell w}{s^2} \ .
    \label{eqs:pp}
\end{align}
Participation ratios are written with first the dielectric factor, then the dielectric thickness, and finally the geometric factors for the design.

\section{Thickness correction}

The finite thickness of the metal film changes the surface electric fields mostly at the edges of the film.  Since the edge fields will be similar for different geometries, their effect will be calculated here for a simple flat coaxial film.  The resulting simple correction to the surface energy can then be applied to different geometries.  

\begin{figure}[t]
\includegraphics[width=0.48\textwidth, 
trim = 140 10 160 20,clip]
{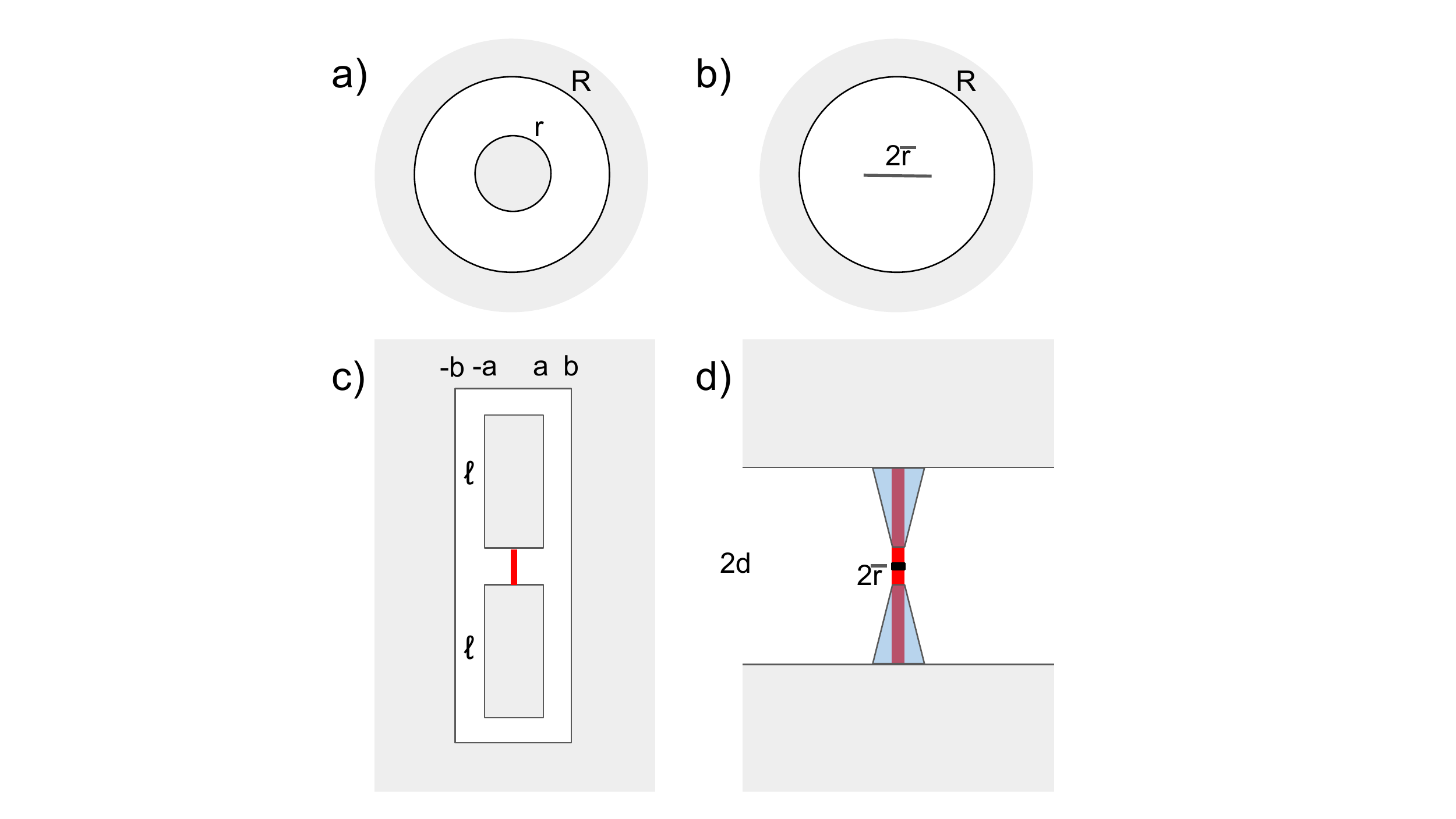}
\caption{\textbf{Design geometries.}
Drawings of capacitor geometries considered here.  a) Cross section of coax, with inner radius $r$ and outer radius $R$.  b) Cross section of flat coax, with width $2\rb$ of inner thin film.  c) Top view of coplanar capacitor, with width $2a$ and length $\ell$ for each section, separated from the ground plane by gap $b-a$.  d) Enlarged view of junction wiring, with length $d$ for each wire from junction to capacitor film.  The width of the wire is $2\rb$ for a straight wire, drawn in red.  A tapered wire is drawn in blue.  
}
\label{fig:Types}
\end{figure}

It is useful to start with a 2-D solution of a coax line, with an inner conductor of radius $r$ and an outer conductor of radius $R$ as illustrated in Fig.\,\ref{fig:Types}a.  The solution for the radial electric field on the surface of the inner conductor is
\begin{align}
E_\textrm{c} & = \frac{V}{r \ln(R/r)}, \label{Ecoax}
\end{align}
with its strength decreasing with radius $x$ as
\begin{align}
E_\textrm{c}(x) = (r/x) E_\textrm{c}  \label{Ecoax_x}
\end{align}

The electric field energy $(\epsilon/2) \int E^2 dv$ is calculated from a volume integral $dv$ of the electric field $E$.  Because the interest here is for the surface energy in a 2-D geometry, we compute the surface energy $U/\ell$ for a line length $\ell$ so that the full energy will be multiplied by the surface thickness and length $\ell$.  For the coax geometry, the surface energy of the inner metal at radius $r$ is
\begin{align}
    U_\textrm{c}^\textrm{m}/\ell & = 
    (\epsilon/2) 2 \pi r E_\textrm{c}^2 \\
    & = \epsilon E_c^2  r\, \pi . \label{Ucoax}
\end{align}
The surface energy corresponding to a substrate surface along a cut through the middle of the coax is
\begin{align}
    U_\textrm{c}^\textrm{s}/\ell & = 
    (\epsilon/2)\ 2 \int_r^R E_\textrm{c}(x)^2 dx \\
    & = \epsilon E_c^2 r\, [1-r/R] ,
\end{align}
where the factor of 2 before the integral comes from the left and right substrate sides. 

Fig.\,\ref{fig:Types}b shows a flat coax, where the circular inner conductor is replaced by a thin film of width $2\rb$.  The electric field magnitude along the coordinate $x$ for this flat is found from numerical solutions for all $x$ to be given by a conformal-mapping solution
\begin{align}
    E_\textrm{f} & = \frac{V}{\rb \ln(2R/\rb)} \label{Eflat} \ ,\\
    E_\textrm{f}(x) & = E_\textrm{f} 
        \sqrt{\frac{\rb}{|\rb+x|}} \sqrt{\frac{\rb}{|\rb-x|}} \ , \label{Eflat2}
\end{align}
which fits well for $R > 2 \rb $. The electric field is perpendicular to the metal surface but parallel to the substrate surface.  The voltage integral checks properly
\begin{align}
    \int_{\rb}^R E_\textrm{f}(x)\, dx = V [1+O(\rb ^2/R^2)]\ .
\end{align}

\begin{figure}[b]
\includegraphics[width=0.48\textwidth, 
trim = 110 20 150 40,clip]
{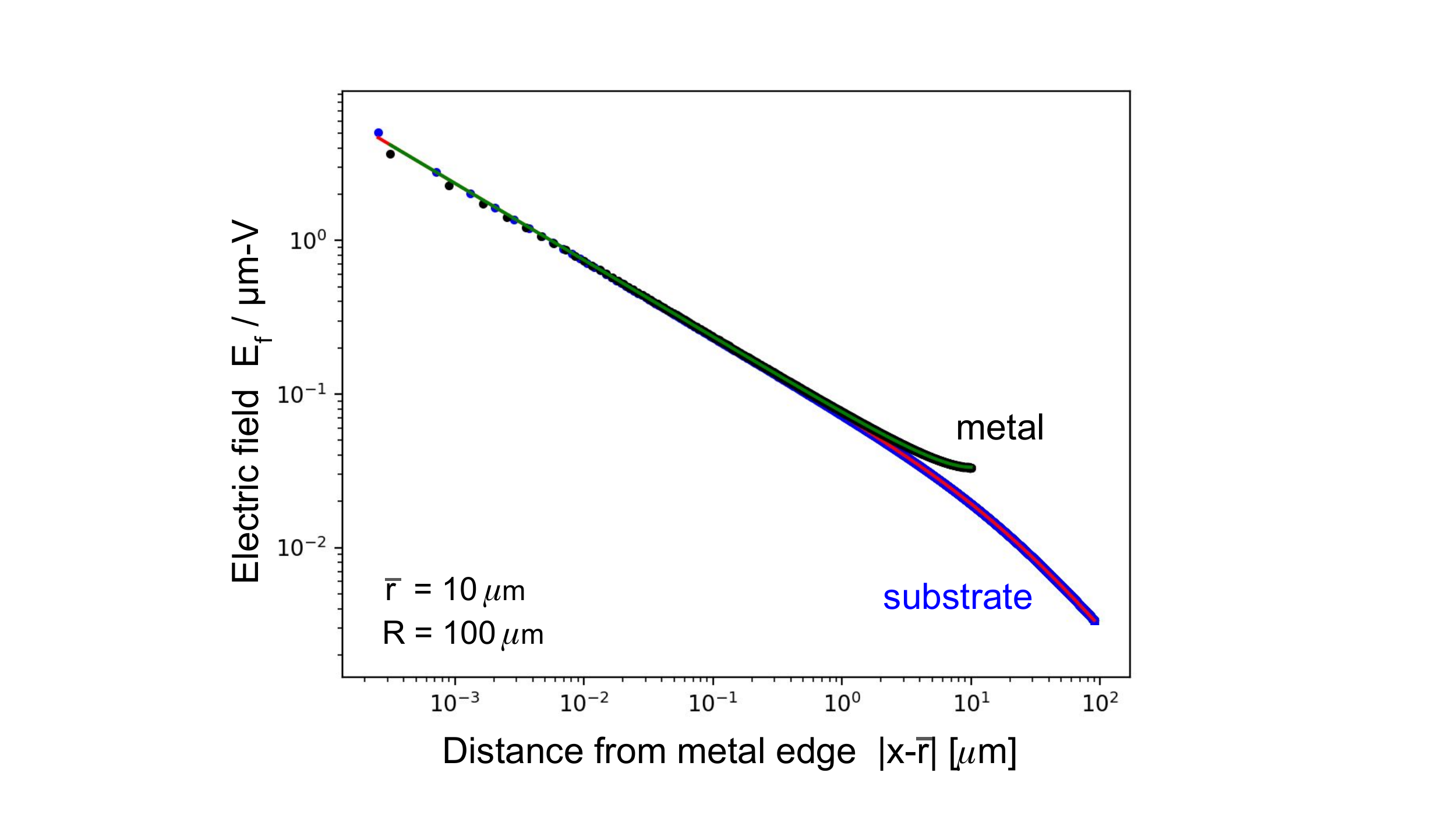}
\caption{\textbf{Electric field of flat coax.}
Plot of the surface electric field for a flat coax for both the metal surface (black) and substrate (blue), obtained by numerical simulation (dots).  The solid lines (green, red) are predictions from Eqs.\,(\ref{Eflat})-(\ref{Eflat2}) and fit well the numerics.  Parameters are $\rb = 10\,\mu\textrm{m}$ and $R = 100\,\mu\textrm{m}$
}
\label{fig:Flat}
\end{figure}

Figure \ref{fig:Flat} shows a comparison between the numerical solution and the formula of Eqs.\,(\ref{Eflat})-(\ref{Eflat2}), showing excellent agreement. The square-root divergence at the metal edge is characteristic of the electric fields of thin metal films.  

The surface energy for the metal surface ($|x| < \rb$) is
\begin{align}
    U_\textrm{f}^\textrm{m1}/\ell & = (\epsilon/2) E_f^2 \ 4\int_0^{\rb-t/2} \frac{\rb^2}{\rb^2-x^2} \ dx \\
    & \simeq \epsilon E_\textrm{f}^2 \rb \ln(4\rb/t) \label{eq:ufm1}
\end{align}
where the factor of 4 is for the top/bottom and left/right parts of the metal, and the logarithmic divergence in the integral at the edge is cut-off at half the thickness $t$ of the film.  

Numerical simulation for a film with a rectangular cross-section of thickness $t$ shows that the electric fields within $t/2$ of the outside corner have a power law behavior with exponent $p = -1/3$, as appropriate for a 90 degree corner \cite{wenner_surfloss,Jackson}.  As an initial approximate solution, this power law dependence of the corner field is then matched to the computed field $E_\textrm{f}(\rb-t/2)$ at a distance $t/2$ from the corner.  At a distance $r_c$ from the corner, the corner field is
\begin{align}
    E_\textrm{c} & = E_\textrm{f}(\rb-t/2) \ [r_c/(t/2)]^{p} 
    \label{eq:edgec} \\
    &= E_f \sqrt{\rb/t} \ [2r_c/t]^{p} .
\end{align}
Including all 4 corners, with 2 sides per corner, the line energy for the corner is approximately
\begin{align}
    U_\textrm{f}^\textrm{m2}/\ell &=  (\epsilon/2) E_f^2 (\rb/t)\ 8
    \int_0^{t/2} [2r_c/t]^{2p} dr_c. \\
    & = 4 \epsilon E_f^2 (\rb/t) (t/2)/(1+2p) \\
    & = \epsilon E_f^2 \rb \ [2/ (1+2p)] \ . \label{Ufm2}
\end{align}
With $1+2p = 1/3$, the numerical factor in Eq.\,(\ref{Ufm2}) is 6 and does not depend on $t$. 

\begin{figure}[t]
\includegraphics[width=0.48\textwidth, 
trim = 140 45 180 55,clip]
{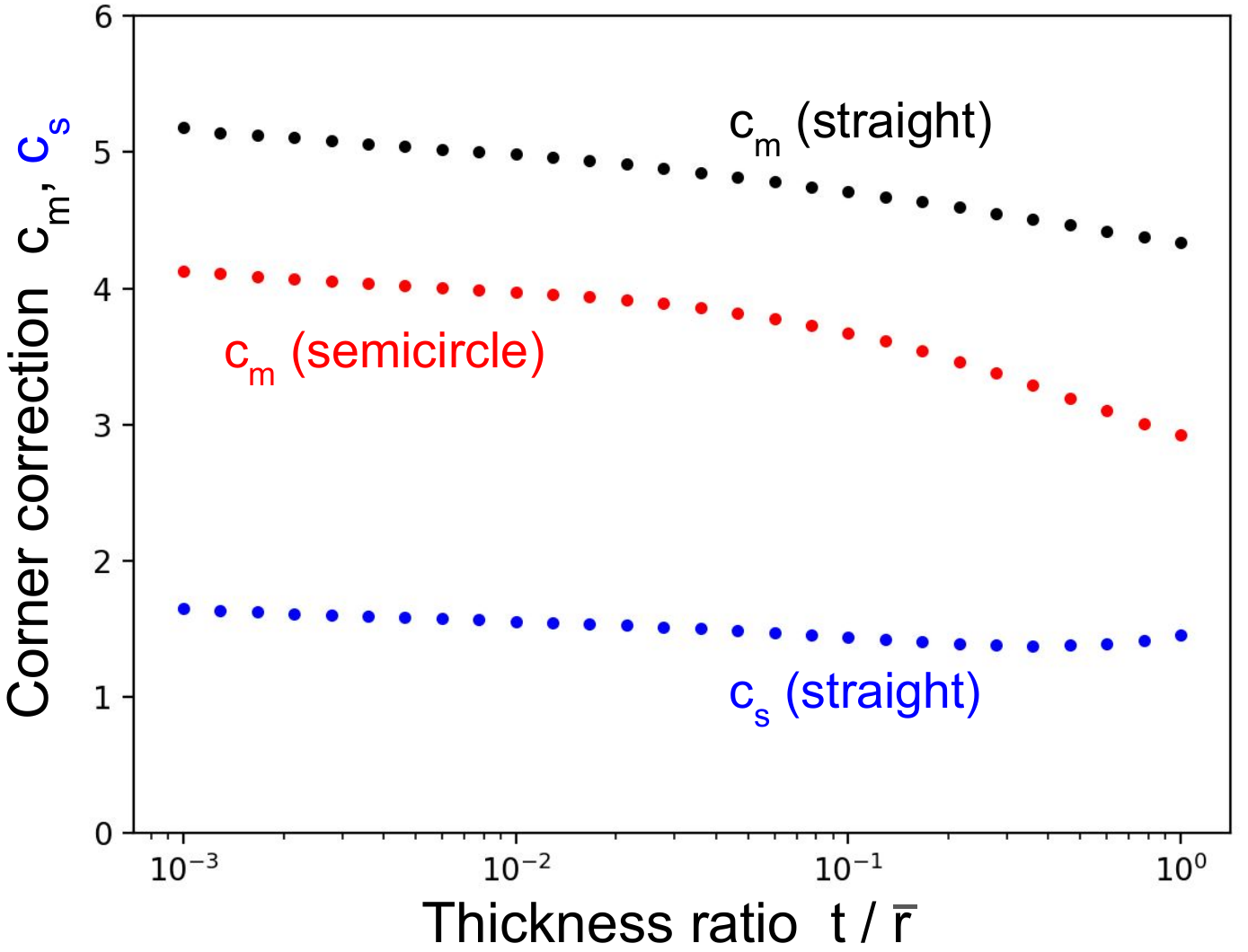}
\caption{\textbf{Corner corrections.}
Plot of corner corrections obtained by numerical simulation of films with finite thickness $t/\rb$ and a straight vertical edge, for the metal $c_m$ (black) and substrate $c_s$ (blue).  Data in red is for a semi-circular edge, showing that sharp corners have a non-negligible but non-dominant effect.  The corrections vary slowly with thickness and are taken as $c_m = 5.0$ and $c_s = 1.6$
}
\label{fig:Corner}
\end{figure}

The total surface energy for the metal is the sum of the two energies
\begin{align}
    U_\textrm{f}^\textrm{m}/\ell & = \epsilon E_\textrm{f}^2 \rb \ 
    [\ln(4\rb/t) + c_m] \label{eq:ufm} \ ,
\end{align}
where $c_m$ is the corner correction for the finite thickness of the metal. Figure\,\ref{fig:Corner} gives $c_m$ obtained from numerical integration of the surface energy.  The corner correction is slowly varying with relative film thickness $t/\rb$ and has a typical value
\begin{align}
        c_m &= 5.0
\end{align}
close to the value 6 obtained above by scaling of the corner fields.  It is useful that the surface energy is predicted well even for thick film, with thickness as much as one-half the width.  The edges typically contribute about 1/3 of the total surface energy.  Also shown is the case of a semicircular edge, which lowers the surface energy a non-negligible but small amount, providing a lower bound for the correction of a rounded edge. 

This result shows that a constant term added to the logarithmic cut-off term well represents the corner fields.    Note the similarity to Eq.\,(\ref{Ucoax}).  Here, the bracket term in Eq.\,(\ref{eq:ufm}) is slightly larger than the corresponding $\pi$ constant in Eq.\,(\ref{Ucoax}), as expected since the flat coax has large edge fields.  The correction factor $\ln(4\rb/t) \rightarrow \ln(4\rb/t) + c_m$ will be used in all formulas for the metal edge.

The surface energy for the substrate surface ($\rb < x < R$) is
\begin{align}
    U_\textrm{f}^\textrm{s1}/\ell & = (\epsilon/2) E_f^2 \ 2\int_{\rb+t/2}^R \frac{\rb^2}{x^2-\rb^2} \ dx \\
    & = \epsilon E_\textrm{f}^2\,(\rb/2) \Big[\ln(4\rb/t)-\ln\frac{R+\rb}{R-\rb} \, \Big] \ ,
\end{align}
where the factor of 2 is for the left/right parts of the substrate. Like found for the metal surface, numerical integration for finite thickness gives a corner correction 1.6.  Since typically $\rb \ll R$, the total substrate surface energy is
\begin{align}
    U_\textrm{f}^\textrm{s}/\ell & = \epsilon E_\textrm{f}^2\, (\rb/2) 
    [\ln(4\rb/t) + c_s - 2\rb/R]  \label{eq:ufs} \ ,\\
    c_s &= 1.6 \ .
\end{align}
This is smaller than the surface energy for the metal surface since it does not include a sharp edge.  The correction factor $\ln(4\rb/t) \rightarrow \ln(4\rb/t)+ c_s$ will be used in all formulas for the substrate edge.

\section{Differential Ribbon Capacitor}

Considered next is the capacitance between the two leads of the qubit, modeled as two long and straight ribbons as illustrated in Fig.\,\ref{fig:Qubit}.  Each ribbon has metal spanning a distance $a$ to $b$ from the centerline, with length $\ell \gg b$ and a metal thickness $t$.  A conformal-mapping solution from Ref.\,\cite{IBM} is used for the electric fields.  The ribbon capacitance for a differential voltage $V$ is 
\begin{align}
    C_\textrm{r} &= [(\epsilon_s + 1)/2] \,\epsilon_0 \ell /C_K(a/b) \ ,
    \label{eq:Cr} \\ 
    C_K &= K(a/b)/K'(a/b) \label{eq:CK}\\
    & \simeq (1/\pi)\ln[2(1+\sqrt{a/b})/(1-\sqrt{a/b})] \ , \label{eq:CKa}\\
    K'(k) &= K(\sqrt{1-k^2})
\end{align}
where $K(k)$ is the complete elliptic integral of the first kind.  Equation\,(\ref{eq:CKa}) is an excellent approximation to Eq.\,(\ref{eq:CK}).  The effective dielectric constant has contribution from both the air ($\epsilon_0/2$) and substrate ($\epsilon_s\epsilon_0/2$).

From the conformal-mapping solution Eq.\,(5) of Ref.\,\cite{IBM}, the  surface fields are
\begin{align}
    |E_r(x)|^2 = \Big( \frac{V/2}{K(a/b)} \Big)^2
    \frac{b^2}{|(x^2-a^2)(x^2-b^2)|} \ , \label{eq:Erib}
\end{align}
where the $E$ field is parallel to the surface on the substrate and perpendicular on the metal.  The surface integral is evaluated in three sections: 
\begin{align}
    &\textrm{inner}\ \ \ \ \ \ \ \ \ \ \ \ \ \ 0 < x < a-t/2 \ ,\\
    &\textrm{center}\ \ \ \ \   a+t/2 < x < b-t/2  \ ,\\
    &\textrm{outer}\ \ \  \ \ \ \,  b+t/2 < x < \infty  \ ,
\end{align}
giving
\begin{align}
    S_i &= \int_0^{a-t/2} dx \, \frac{b^2}{|(x^2-a^2)(x^2-b^2)|} \\
    &= \frac{ \frac{1}{a}\ln\frac{a-x}{a+x}+\frac{1}{b}\ln\frac{b-x}{b+x} }
    {2(1-a^2/b^2)} \ \Big|_0^{a-t/2} \\
    &\simeq \frac{ \frac{1}{a}\ln\frac{4a}{t}+\frac{1}{b}\ln\frac{b-a}{b+a} }
    {2(1-a^2/b^2)} \ , \\
    S_c &\simeq \frac{ \frac{1}{a}(\ln\frac{4a}{t}+\ln\frac{b-a}{b+a})+\frac{1}{b}(\ln\frac{4b}{t}+\ln\frac{b-a}{b+a}) }
    {2(1-a^2/b^2)} \ ,\\
    S_o &\simeq \frac{ \frac{1}{a}\ln\frac{b-a}{b+a}+\frac{1}{b}\ln\frac{4b}{t} }
    {2(1-a^2/b^2)} \ .
\end{align}
Note that $S_c = S_i+S_o$.  

The surface energy of the center metal section is
\begin{align}
   U_\textrm{r}^\textrm{m}/\ell &= (\epsilon/2)\  4 \,[V/2K(a/b)]^2 S_c \\
   &= \frac{\epsilon V^2}{2\,K^2(a/b)} \frac{S_a(c_m)}{a} \label{eq:upm} \ ,\\
    S_a(c_m) &\equiv 
   \frac{ (\ln\frac{4a}{t}+c_m+\ln\frac{b-a}{b+a})+\frac{a}{b}(\ln\frac{4b}{t}+c_m+\ln\frac{b-a}{b+a}) } {2(1-a^2/b^2)} \ .
\end{align}
where the factor of 4 comes from the two ribbons and the top/bottom surfaces.  The dimensionless surface integral is obtained from $S_a/a = S_c$, along with adding the corner correction $c_m$ for a finite thickness 

The surface energy of the inner and outer substrate sections is
\begin{align}
   U_\textrm{r}^\textrm{s}/\ell &= (\epsilon/2)\  2 \,[V/2K(a/b)]^2
   (S_i+S_0) \\
   &= \frac{\epsilon V^2}{4\,K^2(a/b)} \frac{S_a(c_s)}{a} \label{eq:ups} \ ,
\end{align}
where the factor of 2 is from the two sides of the ribbon.  The substrate surface energy is smaller than the metal by approximately a factor of 2.    

The ribbon capacitor has participation ratios coming from the surface-air, metal-substrate, and substrate air interfaces, calculated using Eqs.\,(\ref{eq:MA})-(\ref{eq:SA}), (\ref{eq:upm}) and (\ref{eq:ups})
\begin{align}
    \label{eqs:ribstart}
    p_\textrm{MA}^\textrm{r} =&  
    \frac{1}{\epsilon_\textrm{MA}} 
    \frac{t_\textrm{MA}}{L} 
    \frac{\ell}{a}\ \frac{S_a(c_m)}{2\,K^2(a/b)} \ ,\\
    p_\textrm{MS}^\textrm{r} =& 
    \frac{\epsilon_s^2}{\epsilon_\textrm{MS}}
    \frac{t_\textrm{MS}}{L} \,
    \frac{\ell}{a}\ \frac{S_a(c_m)}{2\,K^2(a/b)} \ ,\\
    p_\textrm{SA}^\textrm{r} = & \,
    \epsilon_\textrm{SA} \,
    \frac{t_\textrm{SA}}{L} 
    \frac{\ell}{a}\ \frac{S_a(c_s)}{2\,K^2(a/b)} \ .
    \label{eqs:ribend}
\end{align}
The only difference in the 3 participation ratios is the dielectric factors and the small changed from the corner constant.  

\begin{figure}[b]
\includegraphics[width=0.48\textwidth, 
trim = 140 65 210 60, clip]
{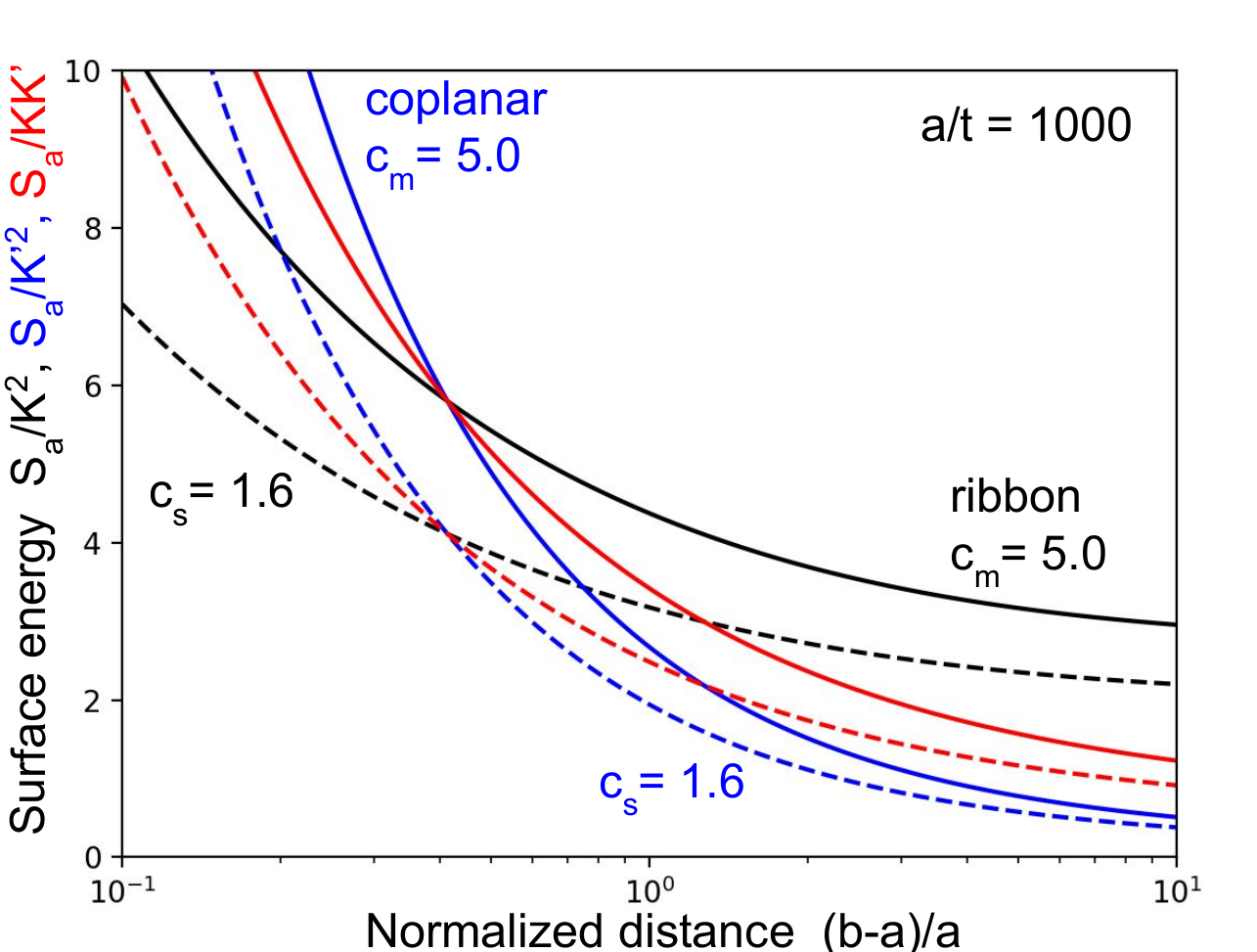}
\caption{\textbf{Surface energies.}
Plot of the normalized surface energies $S_a(c_m)/K^2$ for ribbon (black) and  $S_a(c_s)/K'^2$ for coplanar (blue) geometries, versus the normalized distance $(b-a)/a$.  The plot uses the normalized thickness $a/t = 1000$.  The solid lines are for the metal surface with $c_m = 5.0$, whereas dashes are for the substrate with $c_s = 1.6$.  Also plotted in red is $S_a/KK'$ for the case of all capacitance coming from the ribbon or coplanar geometry.
}
\label{fig:Surf}
\end{figure}

The black lines in Fig.\,\ref{fig:Surf} is a plot of the dimensionless surface energy $S_a(c_m)/K^2$ for the metal (solid) and $S_a(c_s)/K'^2$ for the substrate (dashed) as a function of the normalized distance $(b-a)/a$.  The metal surface energy is greater because of the higher corner constant $c_m > c_s$. As the distance $b-a$ increases, the surface energy decreases.  Typical designs use $(b-a)/a \sim 1$.  Note that the surface energy drops by a non-negligible amount with the lower corner constant, showing that the edge fields from the finite thickness are important.

For the case where all of the capacitance comes from the ribbon $C_r = \epsilon_0 L$, the participation for the metal-substrate interface is
\begin{align}
        p_\textrm{MS}^\textrm{r}(C_r) =& 
    \frac{\epsilon_s^2}{\epsilon_\textrm{MS}(\epsilon_s+1)/2}
    \frac{t_\textrm{MS}}{a} \,\frac{1}{2}
    \ \frac{S_a(c_m)}{K(a/b)K'(a/b)} \ .
\end{align}
The last factor is the geometric mean of the ribbon and coplanar curves of Fig.\,\ref{fig:Surf}, shown in red.  

\section{Differential Coplanar Capacitor}

The qubit can also have capacitance to ground.  This can be modeled as a coplanar structure as shown in Fig.\,\ref{fig:Types}c, where each side of the qubit has a pad with width $2a$ and length $\ell \gg a$, with a ground plane at a distance $b$ from the centerline.  As this is the ``dual'' of the ribbon capacitor, with metal and substrate switched, similar conformal solutions can be used with minor modifications.  The differential capacitance is
\begin{align}
    C_\textrm{c} =& (1/2)[(\epsilon_s + 1)/2]\,\epsilon_0\ell\,4 C_K(a/b) \ , \label{eq:Ccop}
\end{align}
where $C_K$ is defined in Eq.\,(\ref{eq:CK}), and the initial factor of 1/2 comes from the two coplanar capacitors in series.  From Eq.\,(25) of Ref.\,\cite{IBM}, the electric field for each coplanar capacitor is 
\begin{align}
    |E_c(x)|^2 = \Big( \frac{V/2}{K'(a/b)} \Big)^2
    \frac{b^2}{|(x^2-a^2)(x^2-b^2)|} \ , \label{eq:Ecopl}
\end{align}
where now the field is perpendicular to the substrate in the inner and outer sections, and parallel in the center.  

The surface energy of the metal sections is similar to the ribbon case except for an extra factor of 2 to account for the series capacitors, as can be seen from Fig.\,\ref{fig:Types}c since $\ell$ only accounts for half of the total length.  The metal and substrate surface energies are 
\begin{align}
   U_\textrm{c}^\textrm{m}/\ell 
   &= \frac{\epsilon V^2}{K'^2(a/b)} \frac{S_a(c_m)}{a} \label{eq:ucm} \ , \\
   U_\textrm{c}^\textrm{s}/\ell 
   &= \frac{\epsilon V^2}{2\,K'^2(a/b)} \frac{S_a(c_s)}{a} \label{eq:ucs} \ ,
\end{align}
Note the similarities to the ribbon formulas.  The participation ratios are twice as large as the ribbon and with $K$ replaced by $K'$
\begin{align}
    \label{eqs:copstart}
    p_\textrm{MA}^\textrm{c} =&  
    \frac{1}{\epsilon_\textrm{MA}} 
    \frac{t_\textrm{MA}}{L} 
    \frac{2\,\ell}{a}\ \frac{S_a(c_m)}{2\,K'^2(a/b)} \ ,\\
    p_\textrm{MS}^\textrm{c} =& 
    \frac{\epsilon_s^2}{\epsilon_\textrm{MS}}
    \frac{t_\textrm{MS}}{L} \,
    \frac{2\,\ell}{a}\ \frac{S_a(c_m)}{2\,K'^2(a/b)} \ ,\\
    p_\textrm{SA}^\textrm{c} = & \,
    \epsilon_\textrm{SA} \,
    \frac{t_\textrm{SA}}{L} 
    \frac{2\,\ell}{a}\ \frac{S_a(c_s)}{2\,K'^2(a/b)} \ .
    \label{eqs:copend}
\end{align}

For the case where all the capacitance comes from the coplanar structure, the participation is the same as the ribbon design, for example
\begin{align}
        p_\textrm{MS}^\textrm{c}(C_c) = 
        p_\textrm{MS}^\textrm{r}(C_r) \ . \label{eq:Ccoppart}
\end{align}

A single-ended coplanar design is used to test resonators.  In this case, the coplanar capacitance of Eq.\,(\ref{eq:Ccop}) does not have the initial 1/2 term.  The surface energy $U$ and participation ratios are a factor of 2 larger.  The participation $p_\textrm{MS}^\textrm{r}(C_c)$ includes these two factors, so Eq.\,(\ref{eq:Ccoppart}) is unchanged for the single-ended design.  

\section{Differential Ribbon Capacitor \\ With Ground}

Planar transmons are typically designed to have a ground plane surrounding the qubit capacitor, as shown in Fig.\,\ref{fig:full}.  For the ribbon capacitor considered previously, a ground plane is included here from $c$ to infinity and $-c$ to minus infinity, where $c>b$.  The capacitance and surface loss is computed numerically and then fit to functions based on the previous ribbon formulas.  

\begin{figure}[t]
\includegraphics[width=0.48\textwidth, 
trim = 170 20 220 30,clip]
{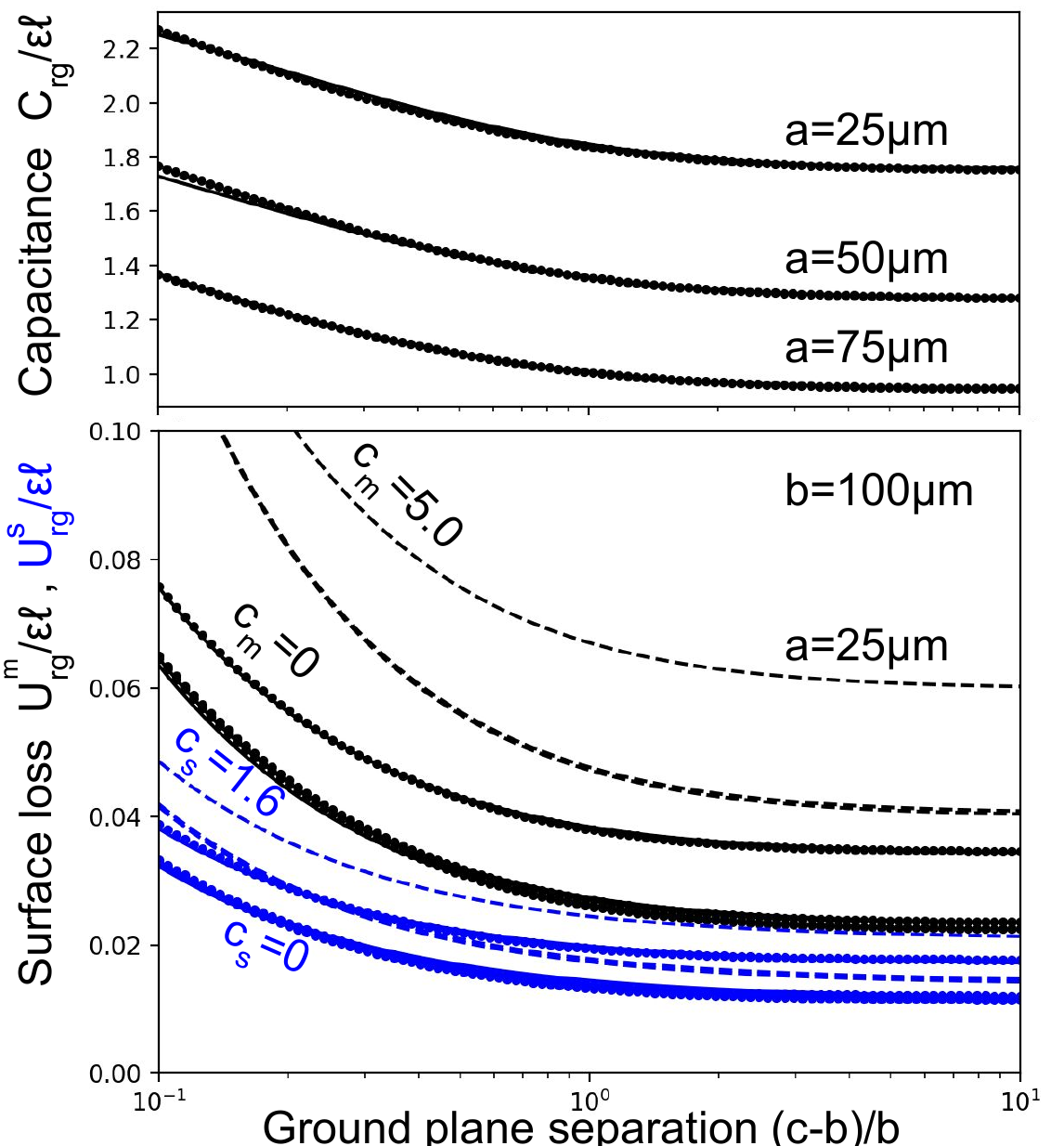}
\caption{\textbf{Ribbon capacitor with ground.}
Plot of the capacitance (top panel) and surface loss (bottom panel) versus the ground plane separation $(c-b)/b$, for $a = (25,50,70)\,\mu$m (top to bottom) and $b = 100\,\mu$m.  Points are numerical simulations and lines are fit formulas.  The metal and substrate surface loss is colored black and blue, respectively.  Numerical simulations are for a infinitely thin metal with an integration cutoff $t/2=0.05\,\mu$m, with the fit also using $c_m=c_s=0$.  Dashed lines include corner correction $c_m=5.0$ and $c_s=1.6$.
}
\label{fig:RibGnd}
\end{figure}

The surface electric field is well described a simple modification to the ribbon case of Eq.\,(\ref{eq:Erib})
\begin{align}
    |E_{rg}(x)|^2 = |E_r(x)|^2 
    \frac{c^2}{|x^2-c^2|} \ . \label{eq:Eribgnd}
\end{align}
Integration of surface charge from the numerical solutions gives a simple modification to the ribbon differential capacitance of Eq.\,(\ref{eq:Cr}) 
\begin{align}
    C_\textrm{rg} &\simeq C_\textrm{r}/[1-(x_e/c)^2]^{0.23} \ ,\\
    x_e &= b -0.15(b-1.2a)  \ .
\end{align}
Figure\,\ref{fig:RibGnd} shows the numerical results (points) and the fit function (line) versus ground plane separation $(c-b)/b$ for three values of $a$, representative of $a/b$ ratios that would commonly be used.  The fit function represents the numerical results well.  

For the surface loss of the metal, the fit function for the numerical results are
\begin{align}
       U_\textrm{rg}^\textrm{m}/\ell &= \epsilon V^2 \Bigg[
   \frac{0.98}{2\,K^2(a/b)} \frac{S_a(a,b,t,c_m)}{a} \\
    & \ \ \ \ \ \ 
    + \frac{1.70}{2\,K'^2(b/c)} \frac{S_{ao}(b,c,t,c_m)}{b} \Bigg] \ ,
\end{align}
where the contribution of $S_{ao}$ corresponds to the outer metal of a coplanar capacitor between $b$ and $c$
\begin{align}
        S_{ao}(b,c,t,c_m) &= \frac{ \ln\frac{c-b}{c+b}+\frac{b}{c}(\ln\frac{4c}{t}+c_m) }{2(1-b^2/c^2)} \ .
\end{align}
The surface loss for both the inner and outer substrate gaps gives the fit function
\begin{align}
    U_\textrm{rg}^\textrm{s}/\ell &= \epsilon V^2 \Bigg[
   \frac{0.95}{4\,K^2(a/b)} \frac{S_a(a,b,t,c_s)}{a} \\
   & \ \ \ \ \ \ 
   + \frac{0.80}{4\,K'^2(b/c)} \frac{S_{a}(b,c,t,c_s)}{b} \Bigg] \ ,
\end{align}
where the second contribution of $S_a$ corresponds to the substrate of a coplanar capacitor between $b$ and $c$. Figure\,\ref{fig:RibGnd} shows the numerical results are well represented by the fit functions. 

\section{Differential Junction Wires}

\begin{figure}[t]
\includegraphics[width=0.48\textwidth, 
trim = 100 20 150 40,clip]
{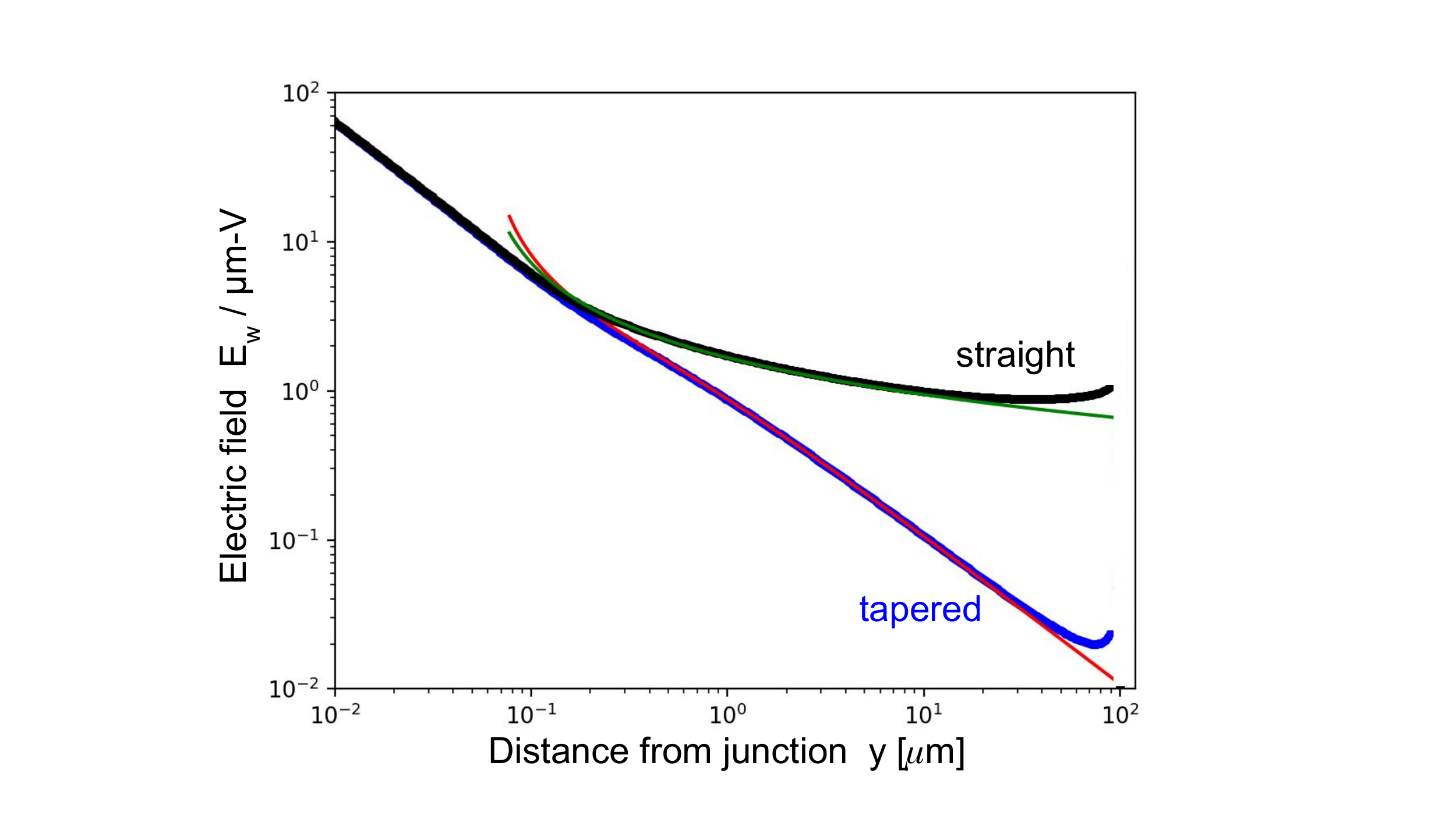}
\caption{\textbf{Electric field of cylindrical wire.}
Plot of the surface electric field for a cylindrical wire for both straight ($S=0$) and tapered ($S=0.2$) radius.  Numerical solutions (dots) match well with the approximation formulas (lines) of Eq.\,(\ref{eq:ewapx}).  Parameters are $r = 0.1\,\mu\textrm{m}$ and $d =  100\,\mu\textrm{m}$.  The uptick of the field at the end of the wire is expected for an edge field.
}
\label{fig:Cyl}
\end{figure}

The connections between the Josephson junction and the capacitor electrodes are made through two junction wires of total length $2d$, as shown in Fig.\,\ref{fig:Types}d.  Treating these wires as round with radius $r$ and placed end-to-end each with length $d$, the surface field can be calculated numerically using the potential matrix Eq.\,(\ref{Mcyl}) for a 2-D geometry with cylindrical symmetry.  For a differential voltage $V$, the surface electric field as a function of distance $y$ from the junction is well described by
\begin{align}
    E_\textrm{cw}(y) = \frac{1}{2}\ \frac{V}{r \ln(2y/r)} \ , \label{eq:ewapx}
\end{align}
as shown in Fig.\,\ref{fig:Cyl}.  In comparison with Eq.\,(\ref{Ecoax}), the second term is equivalent to a coax of inner radius $r$ and outer radius $2y$.  The first term 1/2 represents this coax placed in series with a second coax of the same dimensions, which represents the fields emanating from one circular wire, expanding to a distance $2y$, and then converging in to the other circular wire. 

This 2-D numerical calculation also allows the radius $r$ to change with distance $y$.  Modeling a linear taper with $r = S\,y$, the electric field is found to be well described by Eq.\,(\ref{eq:ewapx}) with $r$ replaced by $r(y)$, as long as the taper is not too large $S < 0.4$;  larger slopes are found not to reduce the electric field significantly.   Figure\,\ref{fig:Cyl} shows numerical results for both a straight and tapered cylindrical wire, with good agreement to the approximation formula Eq.\,(\ref{eq:ewapx}).

A solution for a flat wire can be obtained by assuming the wire has the $x$-dependence of the electric field as given in Eq.\,(\ref{Eflat2}), but with an overall dependence of $E_\textrm{fw}$ with $y$ that is determined numerically.  The appendix shows how to solve this problem with a potential matrix.  Numerical solutions for both straight and tapered flat wires show that a good fitting function is
\begin{align}
    E_\textrm{fw}(y) = \frac{1}{2}\ \frac{V}{\rb\, \ln(4y/\rb)} \ ,
    \label{eq:Efw}
\end{align}
which has the form of $E_\textrm{f}$ in Eq.\,(\ref{Eflat}) but with $2R$ replaced by $4y$.  It is again valid for small slope $S < 0.4$.  The factor of 4 in the logarithm can be understood as a factor of 2 from the coax to the wire geometry, and another factor of 2 from the circular to flat coax formula in Eqs.\,(\ref{Ecoax}) to (\ref{Eflat}).

The metal surface energy for a straight wire of constant width can thus be found by integrating this surface field, which is equivalent to integrating one-fourth of the line energy Eq.\,(\ref{eq:ufm}) over the wire length
\begin{align}
    U_\textrm{sw}^\textrm{m} & =
    2 \int_{2\rb}^d [\tfrac{1}{4} U_\textrm{f}^{m}(R=2y)/\ell]\, dy \\
    &=
    2 \epsilon V^2 \int_{2\rb}^d 
    \frac{\ln(4\rb/t)+c_m}{4\ \rb \ln^2(4y/\rb)}
    \, dy \label{eq:Uswint} \\
    &\simeq \frac{\epsilon V^2}{2} 
    \frac{\ln(4\rb/t)+ c_m}{\ln^2(d/\rb)}\ \frac{d}{\rb}  \ , \label{eq:uwmint}
\end{align}
where the factor of 2 before the integral accounts for both junction wires.  The last formula was fit to numerical integration.  Because of the $d/\rb$ factor, this surface energy can be large, so a more optimal solution is to taper the wire as explained below.  

Similarly from Eq.\,(\ref{eq:ufs}), the surface energy of the substrate for a straight wire of constant width is
\begin{align}
    U_\textrm{sw}^\textrm{s} & =
    2 \int_{2\rb}^d [\tfrac{1}{4}U_\textrm{f}^{s}(R=2y)/\ell]\, dy \\
    &\simeq  \frac{\epsilon V^2}{4}
    \frac{\ln(4\rb/t)+ c_s}{\ln^2(d/\rb)}
    \ \frac{d}{\rb}   \ , \label{eq:uwsint}
\end{align}

The participation ratios for straight wires are 
\begin{align}
    \label{eqs:swstart}
    p_\textrm{MA}^\textrm{sw} =&  
    \frac{1}{\epsilon_\textrm{MA}} 
    \frac{t_\textrm{MA}}{L}\frac{d}{\rb} \, 
    p_\textrm{M}^{\textrm{sw}\prime} \ , \\
    p_\textrm{MS}^\textrm{sw} =& 
    \frac{\epsilon_s^2}{\epsilon_\textrm{MS}}
    \frac{t_\textrm{MS}}{L}\frac{d}{\rb}\, 
    p_\textrm{M}^{\textrm{sw}\prime} \ , \\
    p_\textrm{SA}^\textrm{sw} =& \,
    \epsilon_\textrm{SA} \,
    \frac{t_\textrm{SA}}{L}\frac{d}{\rb} \,
    p_\textrm{S}^{\textrm{sw}\prime} \ , 
\end{align}
where multiplicative factors are given by
\begin{align}
    p_\textrm{M}^{\textrm{sw}\prime} =& 
    \frac{1}{2}\frac{\ln(4\rb/t)+c_m}{ \ln^2(d/\rb)} \ ,\\
    p_\textrm{S}^{\textrm{sw}\prime} =&  
    \frac{1}{2}\frac{\ln(4\rb/t)+c_s}{ \ln^2(d/\rb)} \ .
    \label{eqs:swend}
\end{align}
As found previously, the equations differ only in the epsilon factors and the corner constant.

The capacitance of the straight junction wires is found from numerical simulation
\begin{align}
    C_\textrm{sw} \simeq 
    4.1\, [(\epsilon_s + 1)/2]\,\epsilon_0 \,d/\ln(d/\rb) \ .
\end{align}

\section{Tapered Junction Wires}

The large $d/\rb$ ratio in the above participation ratios contributes to a large surface energy, since the small width of the wires produce large electric fields at its surface.   As surface loss decreases with increasing size, it is natural to increase the width of the wire to lower loss.  A solution to minimize surface energy is to taper the wire, increasing the wire width with increasing distance $y$ from the junction as shown in Fig.\,\ref{fig:Types}d.  The contribution to the line energy, the surface energy per line length $dy$, is the integrand of Eq.\,(\ref{eq:Uswint}), where $\rb$ is now a function of $y$.  The integrand is is minimized at distances $y/t = (10, 100, 1000)$ for a half-width $\rb/y = (0.363, 0.402, 0.425)$, respectively. 

An effective solution is to taper the wire according to $\rb(y) = \textrm{max}(\rb_0, (y-5t)S)$ with the taper starting at $y = 5t$, optimizing the slope $S$ for lowest energy.  Numerical integration of the line energy gives the metal surface energy for a tapered wire that is fit by
\begin{align}
    U_\textrm{tw}^\textrm{m}  & \simeq
    0.68 \,\epsilon V^2 
    \frac{\ln(d/\rb_0)}{S}\ 
    \frac{\ln(4Sd/t)+c_m}{\ln^2(4/S)} \,  
      \ , \label{eq:uwsint}
\end{align}
Although this has a minimum energy at slope $S = 0.45$, it is a broad minimum increasing by 2\% at $S=0.28$ and only 10\% at $S=0.16$.  Note that this formula is similar to Eq.\,(\ref{eq:uwmint}) for a constant width wire, except for the logarithm dependence on $d$.  

Similarly, the substrate surface energy for a tapered wire is fit by
\begin{align}
    U_\textrm{tw}^\textrm{s} & \simeq
    0.29 \,\epsilon V^2 
    \frac{\ln(d/\rb_0)}{S} \ 
    \frac{\ln(4Sd/t)+c_s}{\ln^2(4/S)}\, . \label{eq:uwsint}
\end{align}

\begin{figure}[t]
\includegraphics[width=0.48\textwidth, 
trim = 110 20 150 40,clip]
{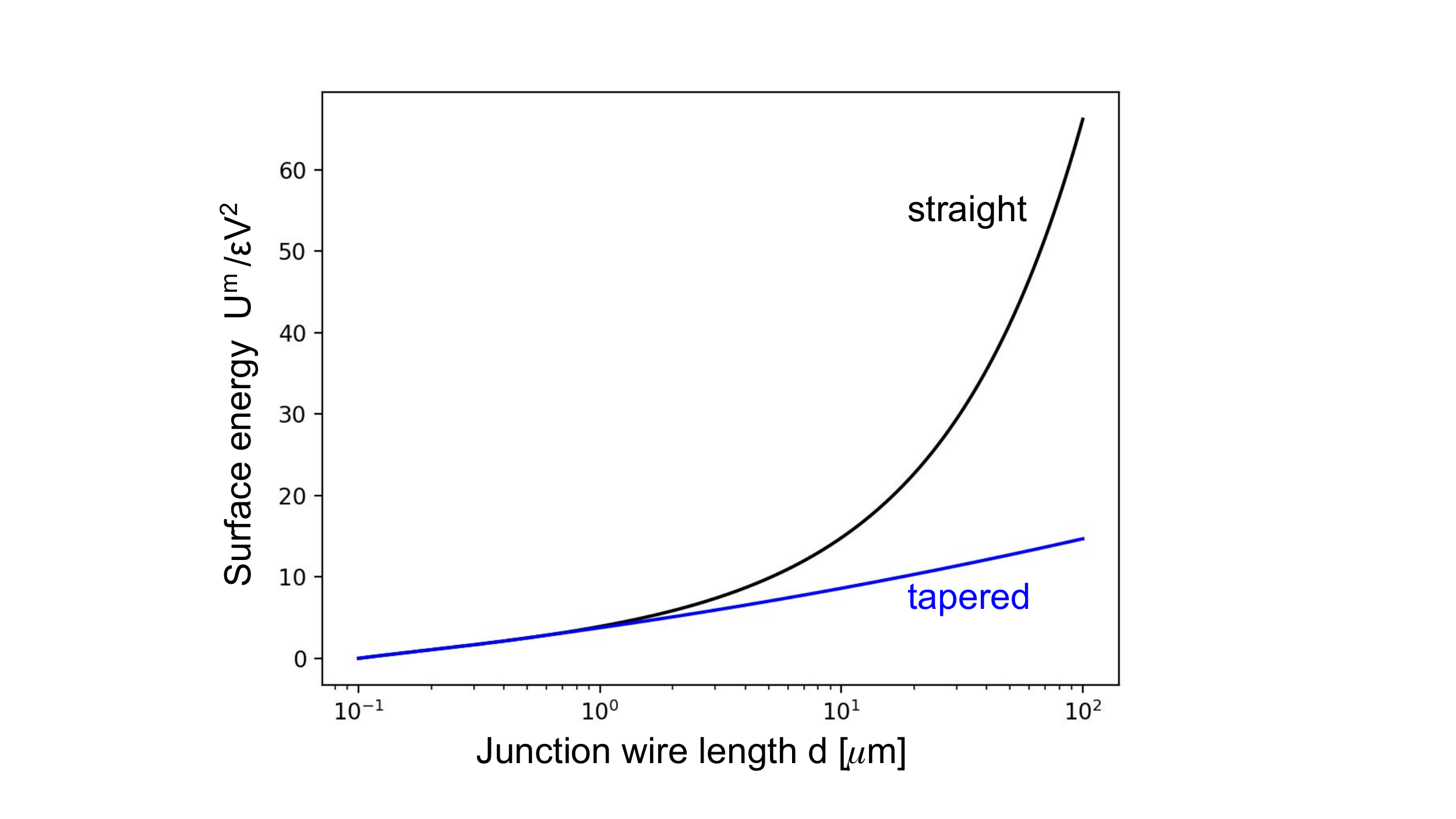}
\caption{\textbf{Junction wire surface energy.}
Metal surface energy (and loss) of the junction wire for straight and tapered designs versus wire length per side $d$.  The tapered wire shows significantly lower energy for wire lengths $d \gtrsim 10\,\mu\textrm{m}$.  At large distances the straight and tapered energy scale with $d$ approximately linearly and logarithmicly, respectively.  Parameters are $t = \rb_0 = 0.1\,\mu\textrm{m}$ and $S=0.4$.  
}
\label{fig:Wires}
\end{figure}

The metal surface loss for the junction wires is plotted in Fig.\,\ref{fig:Wires} for the straight and tapered cases, obtained by numerical integration of Eq.\,(\ref{eq:Uswint}).  At small distances $d \lesssim 5\,\mu\textrm{m}$, the two results are similar, but at large distances the logarithmic scaling makes the tapered loss significantly lower.  It is standard practice to increase the overall size of the qubit capacitor to lower its loss.  When using a large $d$, it is thus increasing important to optimally design the junction wires with a taper.  

The formulas for the participation ratios for a tapered wire are
\begin{align}
    \label{eqs:twstart}
    p_\textrm{MA}^\textrm{tw} =&  
    \frac{1}{\epsilon_\textrm{MA}} 
    \frac{t_\textrm{MA}}{L}\frac{\ln(d/\rb_0)}{S} \, 
    p_\textrm{M}^{\textrm{tw}\prime} \ , \\
    p_\textrm{MS}^\textrm{tw} =& 
    \frac{\epsilon_s^2}{\epsilon_\textrm{MS}}
    \frac{t_\textrm{MS}}{L}\frac{\ln(d/\rb_0)}{S}\, 
    p_\textrm{M}^{\textrm{tw}\prime} \ , \\
    p_\textrm{SA}^\textrm{tw} =& \,
    \epsilon_\textrm{SA} \,
    \frac{t_\textrm{SA}}{L}\frac{\ln(d/\rb_0)}{S} \,
    p_\textrm{S}^{\textrm{tw}\prime} \ , 
\end{align}
with multiplicative factors
\begin{align}
    p_\textrm{M}^{\textrm{tw}\prime} =& 
    0.68 \frac{\ln(4Sd/t)+c_m}{\ln^2(4/S)} \ ,\\
    p_\textrm{S}^{\textrm{tw}\prime} =&  
    0.58 \frac{\ln(4Sd/t)+c_s}{\ln^2(4/S)} \ .
    \label{eqs:twend}
\end{align}

It is recommended using a continuous taper as described above, not a stepped taper, since the continuous taper is optimal at every distance from the junction $x$.  Also,  the sharp corners of the steps will produce large electric fields and increase the surface energy.  

The capacitance of the tapered junction wires is found from numerical simulation
\begin{align}
    C_\textrm{tw} \simeq 
    3.5\, [(\epsilon_s + 1)/2]\,\epsilon_0\, \sqrt{S}\,d \ .
\end{align}

\section{Discussion}

\begin{table}[b]
\begin{tabular}{| l || c | c | c|}
\hline
\textbf{loss$\times$10$^6$ }
& \hspace{8pt} MA \hspace{8pt} 
& \hspace{8pt} MS \hspace{8pt}
& \hspace{8pt} SA \hspace{8pt} \\
\hline
Ref.\,\cite{wenner_surfloss} & 0.10 \hspace{2pt} & 6.13 & 4.02 \\
\hline
This work & 0.060 & 5.93 & 3.57\\
\hline
\hline
\textbf{Participation ratio (\%)} \ \ \ & MA & MS & SA \\
\hline
Ref.\,\cite{partLL} ($d=0.28\,\mu$m) & 0.017 & 0.297 & 0.156 \\
\hline
This work ($d=0$) & \hspace{1.8pt} 0.0012 & 0.139 & 0.027\\
\hline
\end{tabular}
\caption{\textbf{Check data}.  
Comparison of prior numerical results with formulas from this paper.  The first example shows good agreement for $(a,b,t)=(2.5,4.5,0.1)\,\mu$m and surface parameters 
($\epsilon_s, \epsilon_\textrm{MA}, \epsilon_\textrm{MS}, \epsilon_{SA}) = (10,10,10,10)$, thickness 3\,nm and loss tangent 0.002.  
The second example does not agree well and uses $(a,b,t)=(3,6,0.25)\,\mu$m and surface parameters
($\epsilon_s, \epsilon_\textrm{MA}, \epsilon_\textrm{MS}, \epsilon_{SA}) = (11.7,11.4,10,4)$ and thickness 2\,nm, but different trench depths $d$. }
\label{tab:check}
\end{table}

These formulas agree well with a prior numerical simulation, as detailed in Table\,\ref{tab:check}.  The first example shows good agreement with numerical results from Ref.\,\cite{wenner_surfloss}.  Note the only difference in the MS and SA formulas are from the corner constants $c_m$ and $c_s$.  The second example does not agree well with the MS geometry of Ref.\,\cite{partLL}, although the results here are for a flat substrate with no trenching.  It is unexpected that the prior numerical results with trenching gives a higher participation ratio for \textit{all} surfaces.

Surface loss closely scales as the inverse of the system size, as described previously in Ref.\,\cite{wenner_surfloss}.  However, the calculation for the participation ratio from the junction wires have the opposite effect, as its participation increases with length.  Thus there is a crossover in distance $d$ where the surface loss of the wire goes from relatively unimportant to dominant.  Formulas for predicting this crossover is an important result of this work.

Table \ref{tab:ex} shows the participation ratios for the 3 interfaces and 5 qubit capacitance types, for an example geometry of size scale of $\sim 100\,\mu\textrm{m}$ that is appropriate for current devices.  Here a constant thickness 2\,nm of the surface oxides is assumed.  The ribbon has the same participation as the coplanar geometry, as expected.  Of course, predictions depend on actual device parameters, which can be readily made with these formulas. 

For the qubit capacitance, the metal-substrate (MS) interface dominates the surface participation. For the ribbon design, the substrate-air (SA) is about 10 times smaller due to the dielectric factors, half the surface, and a lower corner constant $c_s$.   However, the wire loss is not much smaller and clearly indicates that for present designs this contribution should be carefully considered.  Importantly, the tapering of the wire will produce a significant improvement in qubit performance, about a factor of 2.  

\begin{table}[b]
\begin{tabular}{| c | c || c | c | c|}
\hline
\hspace{15pt} interface \hspace{15pt}
& \hspace{8pt} Eqs. \hspace{8pt}
& \hspace{8pt} MA \hspace{8pt} 
& \hspace{8pt} MS \hspace{8pt}
& \hspace{8pt} SA \hspace{8pt} \\
\hline
\hline
parallel plate & (\ref{eqs:pp})
& 8.16e-5 & & \\
\hline
ribbon & (\ref{eqs:ribstart})-(\ref{eqs:ribend})
& 1.04e-6 & 1.42e-4 & 2.74e-5 \\
\hline
coplanar & (\ref{eqs:copstart})-(\ref{eqs:copend})
& 1.04e-6 & 1.42e-4 & 2.74e-5 \\
\hline
\hline
straight wires & (\ref{eqs:swstart})-(\ref{eqs:swend}) 
& 7.47e-7 & 1.02e-4 & 1.30e-5 \\
\hline
tapered wires & (\ref{eqs:twstart})-(\ref{eqs:twend})
& 4.21e-7 & 5.76e-5 & 9.47e-6 \\
\hline
\end{tabular}
\caption{\textbf{Participation ratios} for various qubit structures.  The top three are for the primary qubit capacitance; for comparison purpose, each uses a length $\ell$ such that its capacitance is 100\,fF.  The bottom two are for straight and tapered wires that connect the junction with the qubit capacitance.  Geometry parameters are 
thickness $t = 0.1\,\,\mu\textrm{m}$; 
parallel plate $(s,w,\ell_p)=(5,100,1130)\,\mu\textrm{m}$;
ribbon and coplanar $(a,b,\ell_r,\ell_c)=(50,100,1391,1138)\,\mu\textrm{m}$;
junction wires $(2d,\rb,\rb_0)=(100,0.1,0.1)\,\mu\textrm{m}$ and $S=0.4$.
Dielectric parameters are 
($\epsilon_s, \epsilon_\textrm{MA}, \epsilon_\textrm{MS}, \epsilon_{SA}) = (11.7,9.8,9.8,3.8)$.  
For simplicity, here the oxide thickness is assumed to be
$t_\textrm{MA} = t_\textrm{MS} = t_\textrm{SA} = 2\,\textrm{nm}$; results can be simply scaled with expected thickness.  
Total loss can be estimated by multiplying the surface loss tangents \cite{partLL}; typical values for amorphous insulators are 0.005 \cite{TLSloss}. }
\label{tab:ex}
\end{table}

When qubit designs use multiple chips that are bump-bonded together, a parallel plate capacitance is often formed between the qubit chip and ground.  Table \ref{tab:ex} shows that the participation ratio of this structure needs to be considered even for a plate separation of $s = 5\,\mu\textrm{m}$, especially since the thickness of the other surfaces are likely less than 3\,nm.   

Although the formulas predict surface energy will decrease slightly with taper slopes greater than 0.4, doing so is not recommended since numerical simulations show that electric fields do not decrease in this range.  Besides, the surface energy only slightly decreases above a slope of 0.2.

An interesting question is how much more surface loss is there for thin films, arising from the large fields at the edges.  It is possible to compare surface energy for a round coax and flat coax of the same width using Eqs.\,(\ref{Ucoax}) and (\ref{eq:ufm}), which shows that the ratio of the metal surface energy is
\begin{align}
   \frac{U_\textrm{f}^\textrm{m}}{U_\textrm{c}^\textrm{m}} & \simeq
   \frac{\ln(4\rb/t)+c_m}{\pi} 
   \simeq 4.0 \ 
\end{align}
for $\rb = 50\,\mu\textrm{m}$ and $t = 0.1\,\mu\textrm{m}$, typical dimensions considered here.  Although the metal-film edges produce more loss, the increase is still acceptable.  Note that the logarithm factor is 7.6, so that about 1/3 of the surface energy comes from the corners within $t/2$ of the edges.

The MA and MS participation formulas in Eqs.\,(\ref{eq:MA})-(\ref{eq:MS}) use surface energies $U/2$ for the two sides of the film, which are then multiplied by dielectric constant factors.  However, since the metal film usually sits on top of the substrate, this splitting of surface energy should change somewhat.  One expects the air side of the surface energy to include both sides of the top corner and the outside of the bottom corner, while the substrate side only includes the film edge of the bottom corner.  Since these two sides of the corner contributes similarly, one expects the constant factor added to the logarithm to be about $1.5\,c_m$ for the air side and and $0.5\,c_m$ for the substrate side.  For example, this modification changes the MA prediction of Table\,\ref{tab:check} from 0.060 to 0.077, closer to the numerical result 0.010.  

Since the MS interface clearly dominates in the participation ratio, there has been effort to minimize this oxide layer by surface treating the silicon wafer before depositing the metal film \cite{HFdip}.  The MS thickness and loss tangent are thus parameters that should be measured carefully to optimize a design.  The qubit capacitor is much larger size than the junction and its wires are often made in a separate step patterned with optical lithography, while the junction and wire is patterned with electron-beam lithography.  If the surface treatment is easier or even possible with the optical lithography step, it is then recommended that the taper is brought down to within $1\,\mu\textrm{m}$ or so of the junction to minimize its loss.  In this case the data in  Fig.\,\ref{fig:Wires} would be used to estimate the loss from both sections of wire; because of the logarithmic dependence, there would still be some contribution from even the short junction section. 

\section{Two-level States}

Surface loss typically comes from two-level states (TLS) \cite{TLSloss}, which saturate and produce less loss at high excitation fields.  Using the numerically computed surface fields, the dependence on power can be found by scaling the reduction in loss from the local electric field $E$ with 
\begin{align}
    E^2 &\rightarrow E^2/\sqrt{1+E^2/E_s^2} \label{eq:TLSsat} \\
    & = E\,E_s \ \ \ \textrm{for}\  E \gg E_s \ ,
\end{align}
where the saturation electric field $E_s$ depends on microscopic parameters of the TLS \cite{TLSloss}.

\begin{figure}[t]
\includegraphics[width=0.48\textwidth, 
trim = 150 10 290 10,clip]
{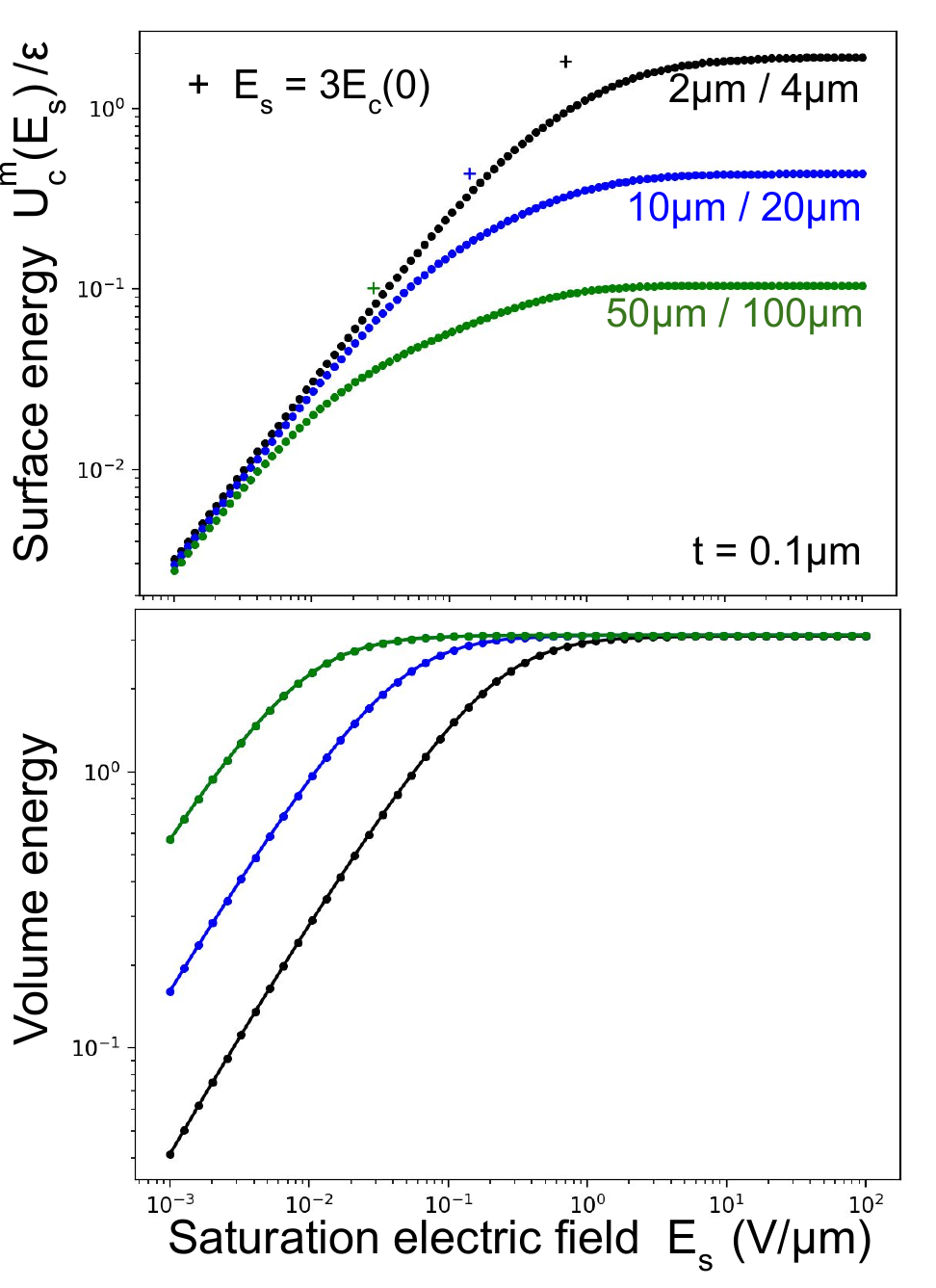}
\caption{\textbf{Saturation loss of a coplanar resonator.}
Top panel: Plot of surface energy versus saturation electric field $E_s$, using surface loss scaled for saturation according to Eq.\,(\ref{eq:TLSsat}).  The single-ended coplanar resonators have voltage $V=1\,$Volt and parameters $a=(2,10,50)\,\mu$m, $b=2a$ and $t=0.1\,\mu$m.  In the low power limit (high saturation $E_s$), the loss is inversely proportional to $1/a$, whereas at high power the curves merge together.  The plus symbol $(+)$ is the characteristic crossover point, given here by the single-ended prediction of the surface energy Eq.\,(\ref{eq:ucm}) and $3E_c(0)$, where $E_c(0)$ is the center $x=0$ electric field of Eq.\,(\ref{eq:Ecopl}).  Bottom panel: Volume energy versus saturation field for the same geometric parameters and colors.  At low fields the volume energy is equivalent to the capacitance per unit length for all curves, as expected.  At left, the characteristic saturation field scales inversely with metal dimension.  
}
\label{fig:Sat}
\end{figure}

Since saturation measurements are typically made with coplanar resonators, numerical integration of the surface loss is shown in Fig.\,\ref{fig:Sat} for three values of $a$, each with the gap equal to the inner metal width $b = 2a$.  As expected, for large saturation fields (loss at low power) the largest resonator gives lowest loss.  At large fields, the loss of all three resonators converges.  This behavior can be understood using dimensional analysis: scaling all the lengths by $D$ decrease the electric field by $E \sim 1/D$, but increases the surface integration by $D$.  For loss at low power, the integral scales as $E^2 D \sim 1/D$.  But when saturated, $E E_s D \sim E_s$ gives constant scaling.  Figure\,\ref{fig:Sat} also shows volume saturation, for example coming from TLS in the substrate.  

The analysis so far has treated the dissipation continuously.  However, surface loss comes from a bath of two-level states, with individual states that are spectroscopically observable for small-area devices \cite{TLSloss}. Simple models predict both the magnitude and the density of TLS, so its spectrum can be extremely useful for identifying the physical location of the loss.  

The dipole moment of the TLS couples to the electric field of the 0 to 1 qubit transition.  This produces a qubit splitting with random frequency and splitting size, but with a maximum splitting size $S_\textrm{max}$ given by Eq.\,(3) of Ref.\,\cite{TLSloss} that is proportional to the qubit electric field and inversely proportional to the square-root of the qubit capacitance.  For a junction capacitor with parallel-plate separation 2\,nm and a qubit capacitance 2\,pF, a value $S_\textrm{max} = $ 74\,MHz was measured.  

\begin{figure}[b]
\includegraphics[width=0.48\textwidth, 
trim = 110 20 180 20,clip]
{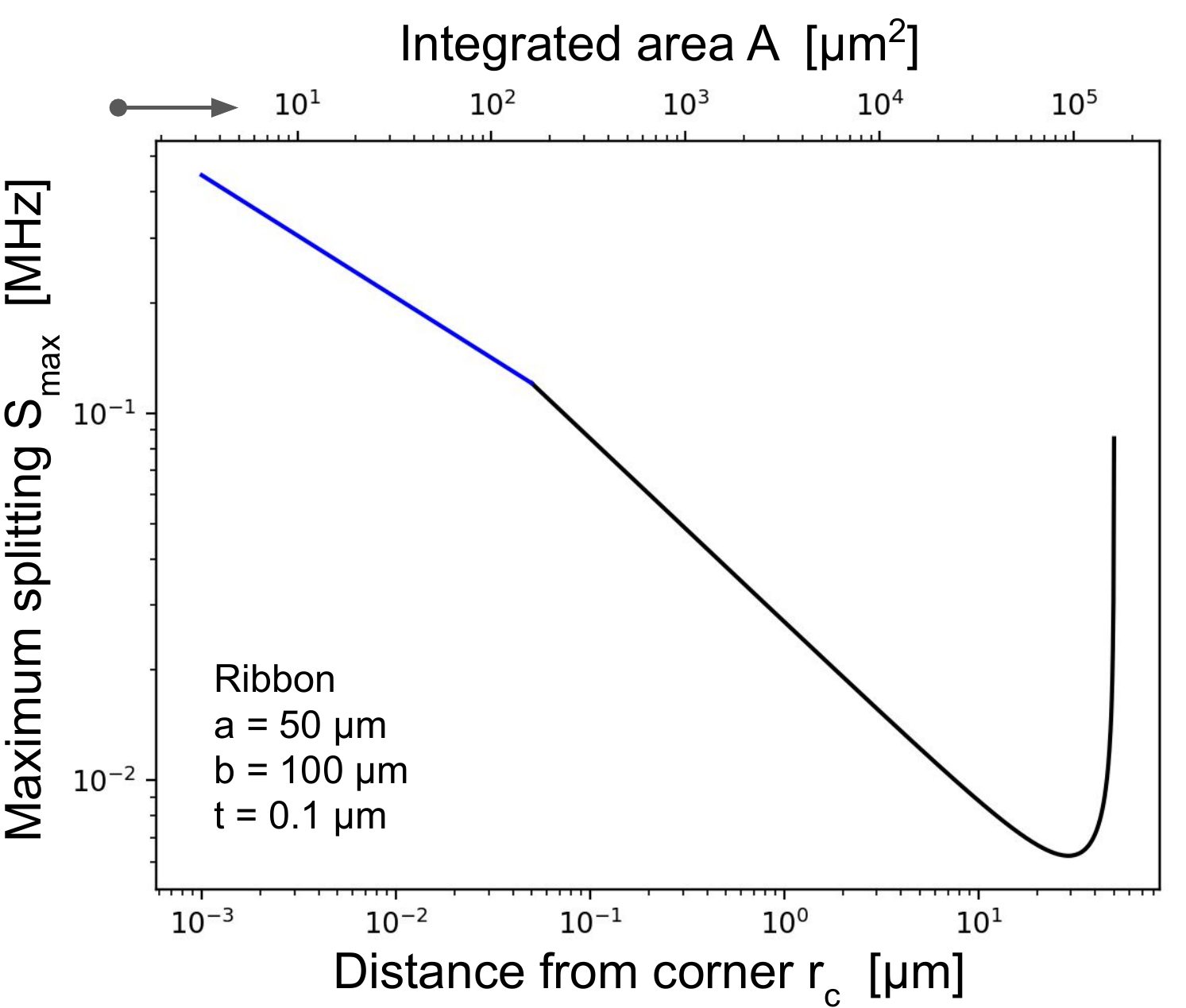}
\caption{\textbf{TLS for ribbon capacitor.}
Plot of maximum splitting $S_\textrm{max}$ versus distance from edge of ribbon capacitor $r_c$.  Black is ribbon solution of $\epsilon_s/\epsilon_{MS}$ multiplied by Eq.\,(\ref{eq:Erib}) ($\sim r_c^{-1/2}$), and blue is edge scaling Eq.\,(\ref{eq:edgec}) ($\sim r_c^{-1/3}$). The integrated area is shown on the top x-axis for $\ell = $1.4\,mm and 2 MS electrodes, showing that the effective areas are typically much greater than $1\,\mu\textrm{m}^2$ (arrow).  
}
\label{fig:TLSrib}
\end{figure}

For a transmon qubit with $C = $ 0.1\,pF, the above scaling gives
\begin{align}
    S_\textrm{max} = (330\,\textrm{MHz} \cdot 2\,\textrm{nm})\ E/V \ ,
\end{align}
where $E/V$ has a dimension of inverse distance and has been computed here for the various surface electric fields.  For the MS interface of a ribbon capacitor, Fig.\,\ref{fig:TLSrib} shows a plot of $S_\textrm{max}$ versus the distance from the inner corner $r_c$, which includes edge corrections at a distance less than the half-thickness $t/2$. The size of the largest splittings are in the few hundred kilohertz range.   

The number of splittings is proportional to the capacitor volume.  Figure\,2 of Ref.\,\cite{TLSloss} shows that the size of the splittings have a log-normal distribution, so that the largest splittings are between $S_\textrm{max}/3$ and $S_\textrm{max}$ and have a density $0.5/\mu\textrm{m}^2$GHz.  The expected TLS density of the ribbon capacitor can be estimated by the effective integrated area $A(S)$, obtained by multiplying $r_c$ by twice the length of the ribbon 2.8\,mm and a factor 3/2 to account for the thicker 3\,nm thick surface oxide.  Since the observed splittings are dominated by those close to $S_\textrm{max}$, the splitting density $\rho_S$ between splittings $S_1$ and $S_2$ is thus approximately given by
\begin{align}
    \rho_S \simeq (0.5/\mu\textrm{m}^2\textrm{GHz})[A(S_2)-A(S_1)] \ .
\end{align}
If one assumes the qubit splitting measurements spans a 2\,GHz frequency range, then the first observable splitting should occur on average for an integrated area $A = 1\,\mu\textrm{m}^2$.  Using Fig.\,\ref{fig:TLSrib}, the largest splittings at 3 nm should have size 300\,kHz, with an average spacing of about one per 200\,MHz in the qubit frequency.  

For a parallel plate capacitor, the effective distance for the electric field is the separation multiplied by $\epsilon_\textrm{MA} = 9.8$.  For the example of Table\,\ref{tab:ex}, this gives $49\,\mu\textrm{m}$.  One finds a splitting size of 13\,kHz, and an effective area of 1.5 times the capacitor area.

\begin{figure}[t]
\includegraphics[width=0.48\textwidth, 
trim = 130 40 175 50,clip]
{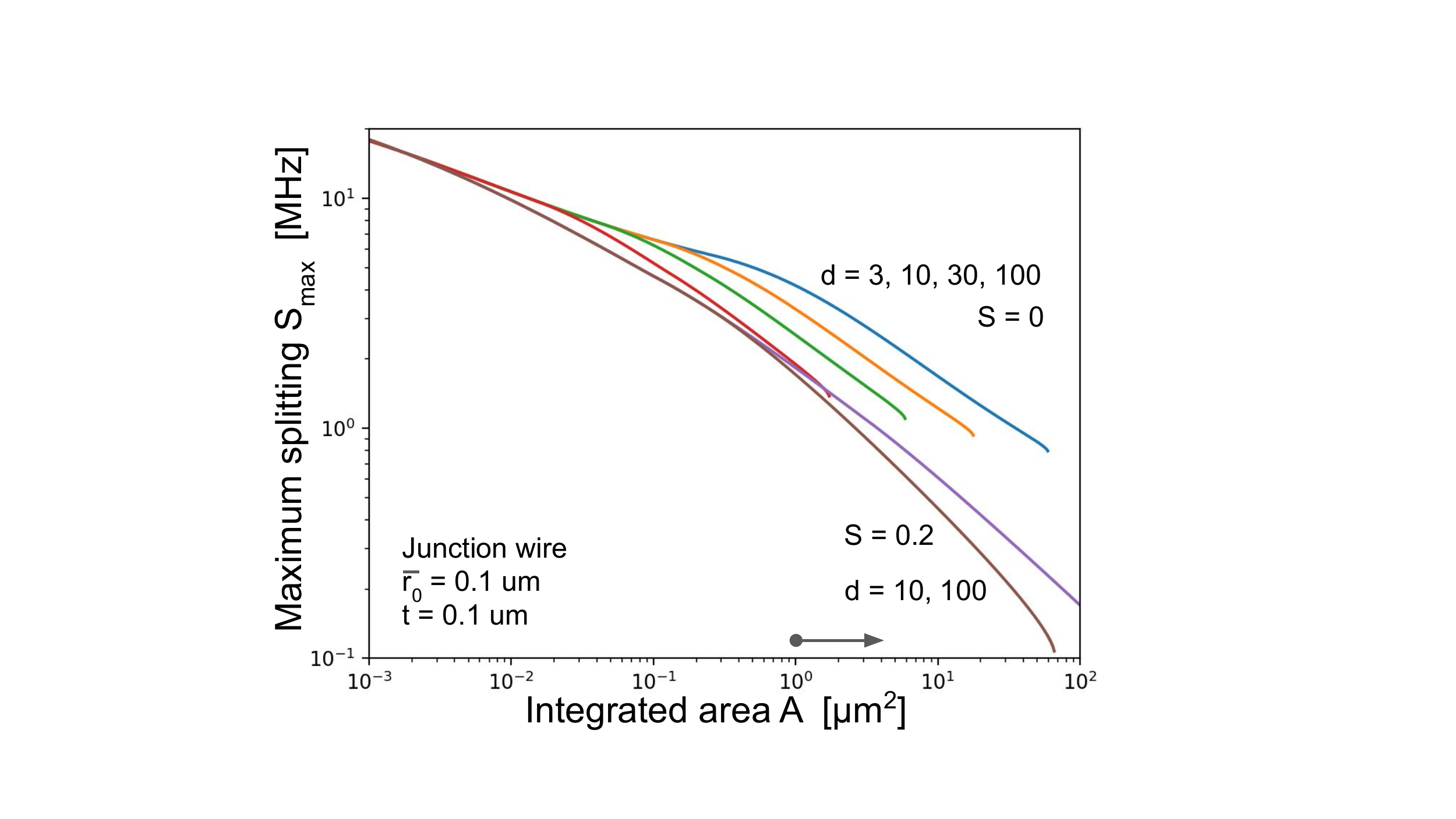}
\caption{\textbf{TLS for junction wire.}
Plot of $S_\textrm{max}$ versus integrated area for untapered ($S = 0$) and tapered ($S=0.2$) wires.  The electric fields are obtained from Eqs.\,(\ref{Eflat2}) and (\ref{eq:Efw}), and numerically breaking up the wire into small area sections.  The tapered wire shows lower maximum splitting, consistent with the continuum results.  For the untapered wire, the observable areas $A > 1\,\mu\textrm{m}^2$ (arrow) has $S_\textrm{max}$ in the 1-4\,MHz range, whereas for tapered it is below 1.5\,MHz.  Note that the dominant contribution comes for wire surfaces at a distance greater than about $10\,\mu\textrm{m}$.   
}
\label{fig:TLSwire}
\end{figure}

Junction wire results are shown in Fig.\,\ref{fig:TLSwire} for the untapered and tapered cases.  These plots were obtained by numerically breaking up the wire into about 100k sections, then computing $S_\textrm{max}$ and the differential area $dA$ for each section.  The curve is obtained by sorting $S_\textrm{max}$ from large to small, and then cumulative summing over the corresponding $dA$ to obtain the integrated area $A$.  The tapered case shows lower splittings $S_\textrm{max}$, consistent with the continuum theory.  The TLS become statistically observable for $A > 1\,\mu\textrm{m}^2$, which 
predicts splittings in the several MHz range.  The dependence on $d$ shows that the dominant contribution to the TLS are for distances greater than about $10\,\mu\textrm{m}$; shorter distances are unimportant because they have small areas. The dominance of an intermediate length scale is perhaps a surprising result, and shows why detailed theory is needed to optimize the wire design.

This result suggests that undercutting the junction wires into the substrate can be an effective solution to decrease the contribution of the metal-substrate interface.  Since there is little contribution at small distances, it is not essential to undercut around the junction, which should improve the reliability of the fabrication and the stability of the junctions.  

\section{Summary}

Calculation of participation ratios and surface loss is challenging because of the divergence of the electric fields at metal edges.  Previously, these fields were solved in the infinitely thin limit using solutions from conformal mapping.  Here, the solutions were extended to the useful limit where the thin surface oxide (few nm) is less than the metal film thickness ($0.1\,\mu$m), and less than the typical film size ($100\,\mu$m).  The finite thickness condition was solved via a calculation that matched the conformal fields to edge fields, then checked and refined with numerical simulation.  Going forward, these formulas are also useful when checking numerical simulations for systematic errors due to meshing.  

Formulas are given for common capacitor structures.  By separating out the geometery of actual designs, participation ratios can be calculated accurately and then used to optimize the design.   This is an important check on numerical calculations since misleading results can come from finite meshing when structures range in size from nanometers to millimeters.

For junction wires, a solution for the capacitance and surface loss was obtained using well formed models, approximations and numerics, which should give accurate and reliable formulas.  A tapered junction wire was shown to have superior performance compared to straight wires when the wire length is longer than about $\sim 10\,\mu$m.  This design feature is important for the latest generation of devices that use large capacitor size to lower surface loss.  A further design improvement for the taper was suggested.

These electric field solutions enable a prediction of the TLS spectrum, which could be invaluable to identify where the TLS comes from in the qubit design.  

Finally, it is hoped that these results will encourage researchers to precisely test surface loss theory, and measure in additional experiments the various surface loss parameters.  By doing so, this should speed the optimization and development of long coherence time qubits.  

\appendix*
\section{Appendix \\ Numerical Calculation of Surface Electric Fields}

For thin films suitable for superconducting qubits, it is useful to numerically calculate the electric fields, for example for a thin film of finite thickness.  Fortunately, realistic transmon designs are well approximated by simple geometries with fields that can be well described using simple fitting functions, so that they can be physically understood and optimized.  Two-dimensional geometries are particularly amenable to efficient numerical solutions and thus their method of solution will be described here first.  For simplicity, the calculations here will assume a constant dielectric constant $\epsilon$.  Corrections due to the substrate and vacuum are included in the main text.  

Figure \ref{fig:Types} shows geometries to be considered here.  The first is 2-dimensional, with solutions given per unit length in the third dimension, with results typically scaled with length $\ell$.  The second solution uses cylindrical symmetry to turn a 3-dimensional problem to 1-D.  The last uses an approximation to the edge fields so that a thin film wire can be similarly calculated in 1-D.

Numerical solutions can be obtained through inverting a matrix.  For a 2-D geometry with translational invariance in the z direction, the problem is first broken into an vector of points in the x-y plane that have line charge $\vec{q}$.  The voltage $\vec{V}$ can be solved with the matrix equation $\vec{V} = \bar{M}\vec{q}$, where the potential matrix $\bar{M}$ has elements
\begin{align}
    M_{ij} & = \frac{1}{2 \pi \epsilon} \ln(1/\rho_{ij}) \ ,
\end{align}
where $\rho_{ij}$ is the distance between point $i$ and $j$.  For metal electrodes the voltage is set instead, so the charge can be obtained using $\bar{q}= \bar{M}^{-1} \vec{V}$, where the inverse matrix $\bar{M}^{-1}$ can be thought of as a capacitance matrix.  The time to solution grows as the cube of the number of points, which can be solved quickly for size 1k - 10k.

For a 3-D geometry with cylindrical symmetry, the potential matrix can be solved for a circular ring of total charge $q$, giving a potential matrix with elements
\begin{align}
    M_{ij} & = \frac{1}{2\pi^2\epsilon}
    \frac{\textrm{ellipk}(-4r_i r_j/\rho_{ij}^2)}{\rho_{ij}}
    \ , \label{Mcyl} \\
        &= \frac{1}{4\pi \epsilon \rho_{ij}} \ \ \textrm{for} \ \ \rho_{ij} \gg r_ir_j \ ,
\end{align}
where $\rho_{ij}$ is the distance between points in the r-z plane, and $r_i$ and $r_j$ are the radial components.  The Python ellipk($m$) function is equivalent to $K(k)$ but with $m = k^2$, and allows negative $m$ arguments.

For a similar 3-D geometry of a flat coax, the potential matrix for the inner conductor is
\begin{align}
        M_{ij} & = \int_{-\rb_j}^{\rb_j} \frac{1/\pi}{\sqrt{\rb_j^2-x^2}}
        \frac{1/4\pi\epsilon}{\sqrt{x^2+y_{ij}^2}}\, dx \\
        &=\frac{1}{2\pi^2\epsilon\ }\,
        \frac{\textrm{ellipk}(- \rb_j^2/y_{ij}^2)}{y_{ij}}
        \ , \label{Mflat} \\
    &= \frac{1}{4\pi\epsilon \, y_{ij}} \ \ \textrm{for}\ \  y_{ij} \gg \rb_j \ ,
\end{align}
where $y_{ij}$ is the distance between points $i$ and $j$ on the centerline of the wire.  In comparison with the matrix for the cylinder geometry Eq.\,(\ref{Mcyl}), the difference is the absence of the factor of 4 in the ellipk argument.

\end{document}